\documentclass[aps,prb,twocolumn,superscriptaddress,floatfix]{revtex4-2}
\usepackage{graphicx,graphics}
\usepackage{dcolumn}
\usepackage{amsmath,amssymb,amsfonts,nccmath}
\usepackage{latexsym,verbatim}
\usepackage{bm}
\usepackage{color}
\usepackage{ulem}
\usepackage{soul}
\usepackage{braket}
\usepackage[percent]{overpic}
\usepackage[breaklinks=true,colorlinks,citecolor=blue,linkcolor=blue,urlcolor=blue]{hyperref}
\usepackage{lipsum}
\usepackage{enumitem}
\usepackage[T1]{fontenc}

\usepackage{xprintlen}

\newcommand{\red}[1]{{\color{red} #1}}

\makeatletter
\renewcommand\@make@capt@title[2]{%
 \@ifx@empty\float@link{\@firstofone}{\expandafter\href\expandafter{\float@link}}%
  {\textbf{#1}}\@caption@fignum@sep#2\quad
}
\makeatother

\usepackage[acronym,nomain,nonumberlist,nopostdot]{glossaries}

\newacronym{soc}{SOC}{Spin-Orbit Coupling}
\newacronym{qvsh}{QVSH}{Quantum Valley Spin Hall}
\newacronym{qsh}{QSH}{Quantum Spin Hall}
\newacronym{jj}{JJ}{Josephson Junction}
\newacronym{gjj}{GJJ}{Graphene Josephson Junction}
\newacronym{sde}{SDE}{Superconducting Diode Effect}
\newacronym{jde}{JDE}{Josephson Diode Effect}
\newacronym{abs}{ABS}{Andreev Bound States}
\newacronym{vz}{VZ-SOC}{Valley-Zeeman SOC}
\newacronym{km}{KM-SOC}{Kane-Mele SOC}
\newacronym{rb}{R-SOC}{Rashba SOC}

\begin{document}
%
\title{From Valley Filtering to Superconducting Diode Effect in Spin–Orbit Coupled Graphene Junctions}
\author{F. Bonasera}
\affiliation{Dipartimento di Fisica e Astronomia ``Ettore Majorana'', Universit\`a di Catania, Via S. Sofia 64, I-95123 Catania,~Italy}
\affiliation{INFN, Sez.~Catania, I-95123 Catania,~Italy}
\affiliation{Centro Siciliano di Fisica Nucleare e Struttura della Materia, Via S. Sofia 64, I-95123}
\author{G. A. Falci}
\affiliation{Dipartimento di Fisica e Astronomia ``Ettore Majorana'', Universit\`a di Catania, Via S. Sofia 64, I-95123 Catania,~Italy}
\affiliation{INFN, Sez.~Catania, I-95123 Catania,~Italy}
%
\author{E. Paladino}
\affiliation{Dipartimento di Fisica e Astronomia ``Ettore Majorana'', Universit\`a di Catania, Via S. Sofia 64, I-95123 Catania,~Italy}
\affiliation{INFN, Sez.~Catania, I-95123 Catania,~Italy}
%
\author{F.M.D. Pellegrino}
\affiliation{Dipartimento di Fisica e Astronomia ``Ettore Majorana'', Universit\`a di Catania, Via S. Sofia 64, I-95123 Catania,~Italy}
\affiliation{INFN, Sez.~Catania, I-95123 Catania,~Italy}
\affiliation{Centro Siciliano di Fisica Nucleare e Struttura della Materia, Via S. Sofia 64, I-95123}

\begin{abstract}

We study the transport properties of proximitized graphene, which can acquire a spin-orbit coupling by the proximity effect with a substrate.
We focus on the ballistic and zero temperature limits, making use of a tight-binding procedure based on the {\sc kwant} Python package.
We first find key results on valley-filtering properties and asymmetric edge transport in spin-orbit coupled graphene single junctions, and then move to the analysis of the superconducting transport in a graphene Josephson junction, in the short junction limit.
We study the relative contribution of edge modes for different edge terminations and some degree of edge disorder, and also analyze the magnetic interference pattern that arises when threading the junction with a perpendicular magnetic field.
We find residual supercurrent at high magnetic fluxes, due to the localized nature of transport in the junction, and a strong non-reciprocal transport that leads to a significant Josephson diode effect.


\end{abstract}

\maketitle

\section{Introduction}
\label{sec:intro}

\glsresetall



The development of clean encapsulated graphene platforms has established graphene as a versatile material for hybrid superconducting devices \cite{Dean_2010_a,Mayorov_2011_a,Wang_2013_a}.
Owing to the exceptional interface quality and high contact transparency achievable in modern heterostructures, \glspl*{gjj} can operate in the ballistic regime, support electrostatically tunable supercurrents, and display markedly non-sinusoidal current–phase relations characterized by high transmission channels \cite{Black-Schaffer_PRB_2008,Black-Schaffer_PRB_2010,Calado_2015_a,BenShalom_2016_a,Borzenets_2016_a,English_2016_a,Nanda_2017_a,Allen_2017_a,Pellegrino_2020_a}.
Recently, it was found that when encapsulated with transition-metal dichalcogenides, graphene can acquire a strong \gls*{soc} by proximity effect \cite{Avsar_2014_a,Mendes_2015_a,Gmitra_2015_a,Wang_2015_b,Alsharari_2016_a,Wakamura_2018_a,Khatibi_2022_a,Sun_2023_a,Zollner_2025_a}.
%
%
Due to its semimetallic nature, proximity-induced \gls*{soc} can open valley-dependent band gaps, leading to different phases of the graphene layer.
Indeed, monolayer graphene was the first material predicted to exhibit the topological \gls*{qsh} phase due to the \gls*{km} \cite{Kane_2005_a,Kane_2005_b}, but its intrinsic \gls*{soc} is too small to observe its effects at approachable temperatures \cite{Konschuh_2010_a, Min_2006_a}.
Moreover, a combination of \gls*{rb} and \gls*{vz} was recently predicted to induce a trivial gapped graphene phase that still hosts metallic edge states \cite{Frank_2018_a}.
These edge states were shown to be protected by time-reversal symmetry and consist of spin-polarized pseudohelical and valley-localized modes.
This graphene state was hence dubbed \gls*{qvsh} phase.
The properties of graphene edge states have been the subject of a lot of recent interest \cite{MarmolejoTejada_2018_a,Wang_2021_b,Lu_2021_a,Prudkovskiy_2022_a,Lu_2024_a,Lu_2026_a}, together with the interplay between edge states in general and superconductivity \cite{Hart_2014_a,Pribiag_2015_a,Tkachov_2015_a,Zyuzin_2015_a,Zhu_2017_,Bocquillon_2017_a,Bours_2018_a,Draelos_2018_a,Seredinski_2019_a,Sticlet_2020_a,HaidekkerGalambos_2020_a,Blasi_2023_a,Vignaud_2023_a,Barrier_2024_a,Rout_2024_a,Jang_2025_a}.

An exciting development in superconducting electronics has been the discovery of the \gls*{sde}, where the critical supercurrent becomes non-reciprocal, meaning it differs for forward and reverse biases \cite{Ando_2020_a,Nadeem_2023_a,Hou_2023_a}.
In Josephson junctions, this effect manifests as $I_{\rm c}^+\neq I_{\rm c}^-$ and takes the name of \gls*{jde} \cite{Zhang_2022_a,Davydova_2022_a}.
Such nonreciprocal supercurrents are of great interest for dissipationless electronics \cite{Nadeem_2023_a}.
The \gls*{sde} and the \gls*{jde} have been explored in a variety of systems including supercurrent interferometers \cite{Souto_2022_a}, systems with \gls*{soc} and magnetic interactions \cite{Reynoso_2008_a,Reynoso_2012_a,Baumgartner_2022_a,Baumgartner_2022_b,Yuan_2022_a,Turini_2022_a,Bauriedl_2022_a,dePicoli_2023_a,Costa_2023_a}, systems based on topological materials \cite{Tanaka_2022_a,Wei_2023_a,Lu_2023_a,Fu_2024_a,Wang_2024_a,Nagahama_2025_a}, multi-terminal Josephson junctions \cite{Gupta_2023_a,Zhang_2024_a,Coraiola_2024_a,Correa_2024_a} and others \cite{Wei_2022_a,Lin_2022_a,Hu_2023_a,DiezMerida_2023_a,Debnath_2024_a}.
The typical key ingredients to observe non-reciprocal superconducting transport are broken time-reversal and inversion symmetries \cite{Davydova_2022_a}.
These conditions are often met by considering systems with a \gls*{soc} interaction paired with a Zeeman field \cite{He_2022_a}.
However, large Zeeman fields can severely suppress superconductivity and hinder device function.
Alternatively, in planar \gls*{jj}, a \gls*{jde} can be obtained by combining the orbital effects of a small perpendicular magnetic field with a mirror asymmetry along the width of the junction, as outlined in Ref.~\cite{Chirolli_2025_a}.


In this work, we analyze the transport properties of a \gls*{gjj} in the \gls*{qvsh} phase, in the ballistic and short junction approximations at zero temperature.
Specifically, we extend our previous work \cite{Bonasera_2025_a} on the bulk transport properties of spin-orbit coupled \gls*{gjj} to consider finite-width effect and the contribution of the edge states.
We work mainly with tight-binding numerical calculations using the {\sc{kwant}} Python package \cite{Groth_2014_a}.
In our normal-state analysis, we find that both the graphene helical and pseudohelical edge states act as an efficient valley filter for bulk electrons, which, in turn, translates to a strong asymmetric transport along the zigzag edges of a \gls*{qvsh} junction.
When analyzing the supercurrent flowing through the \gls*{gjj}, we find it is characterized by strong resonances, which are sensitive to the specific edge termination and robust against moderate edge disorder.
The magnetic interference pattern of the junction shows slowly damped periodic oscillations, which is consistent with edge-dominated transport \cite{Beenakker_2015_a} and with previous experimental work~\cite{Wakamura_2020_a}.
We also find that, in the \gls*{qvsh} graphene phase, a small perpendicular magnetic field produces a \gls*{jde} with efficiencies up to 60\%.
The non-reciprocity of the junction is linked to the \gls*{soc} broken mirror symmetry and the asymmetric edge transport across the junction's edges; together, they enable supercurrent rectification via \gls*{soc} and orbital magnetic effects.
Our results extend previous studies of edge-state-based setups \cite{Scharf_2024_a,Du_2025_a} to graphene, thereby providing a material-specific realization of the proposal presented in Ref.~\cite{Chen_2018_b}.

This paper is organized as follows.
In the first part of Sec.~\ref{sec:model}, we introduce the scattering formalism, together with the relative approximations, used to analyze the transport properties of the system.
In the second part of the section, we introduce the tight-binding Hamiltonian used to model the proximitized graphene layer, together with the different kinds of \gls*{soc} interactions and associated graphene phases.
In Sec.~\ref{sec:NormalState_Transport}, we explore two related properties of transport in non-superconducting single junctions between pristine and \gls*{qsh} or \gls*{qvsh} graphene, namely the valley polarization effect and the asymmetric edge transport.
After this, we focus on the \gls*{qvsh} graphene phase and study its superconducting transport properties in a \gls*{gjj}.
In Sec.~\ref{sec:Edge_Supercurrent}, we study the critical current of the \gls*{gjj}, focusing on edge transport inside the energy gap and analyzing its robustness against scalar disorder in the form of edge defects.
In Sec.~\ref{sec:MIP}, we analyze the magnetic interference pattern of the junction when threaded by a perpendicular magnetic field: we analyze the robustness of the critical current at high magnetic field and instances of non-reciprocal transport. Finally, conclusions are drawn in Sec.~\ref{sec:conclusions}.

\section{Model}\label{sec:model}

\begin{figure}[t!]
\centering
\begin{overpic}[width=0.98\columnwidth,trim={0 0cm 0 0cm}]{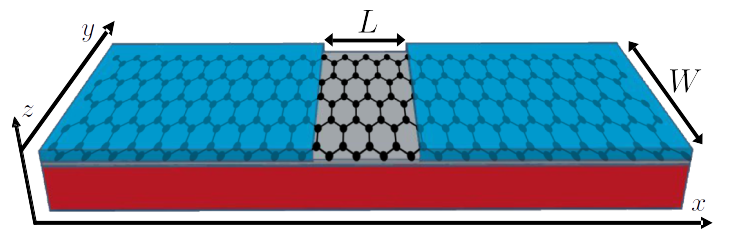}\end{overpic}
\caption{Schematic of the system. A graphene Josephson junction consists of a graphene layer (gray) on a substrate (red), with superconducting electrodes (light blue) covering the $\left|x\right| > L/2$ regions.
}
\label{fig:setup}
\end{figure}

A schematic of the system under study is shown in Fig.~\ref{fig:setup}.
We consider a \gls*{jj} made of a graphene layer (in gray) grown on top of a substrate (in red), which, due to the proximity effect, acquires a \gls*{soc} in the $\left|x\right| < L/2$ region and a superconducting pairing for $\left|x\right| > L/2$ (light blue region).
The width of the junction is finite, $W$, and the edges can have different terminations, such as zigzag, armchair, and in-between ones.
In order to describe the system, we use a tight-binding model with a step-like profile of the superconducting order parameter and \gls*{soc} potentials in the different regions of the junction, 
which is a commonly employed approximation for superconductor–semiconductor junctions \cite{Beenakker_1991_a,Likharev_1979_a}.
The superconducting transport problem can then be solved by focusing on the normal state scattering properties of the inner region of the junction \cite{Beenakker_1991_a}.
In the scattering formalism, the condition for the existence of an \gls*{abs} at energy $\epsilon$ can be expressed as \cite{Beenakker_1991_a,vanHeck_2014_a}
\begin{equation}\label{eq:GeneralABS_Scattering}
    s_{\rm A}(\epsilon) s_{\rm N}(\epsilon) \Psi_{\rm in} = \Psi_{\rm in},
\end{equation}
where $\Psi_{\rm in} = (\Psi_{\rm in}^e,\Psi_{\rm in}^h)$ is a vector with the electron and hole components of a wavefunction incident on the junction,  in the basis of the incoming modes of the normal leads into the scattering region.
Here, $s_{\rm A} (\epsilon)$ is the scattering matrix that links the outgoing modes of the leads to the incoming ones, after reflection with the superconductors.
In a basis where the outgoing modes are the time-reversal symmetric of the incoming ones, $s_{\rm A}(\epsilon)$ can be expressed as
\begin{equation}\label{eq:SA_AndreevApproximation}
    s_{\rm A}(\epsilon) = \alpha(\epsilon) \begin{pmatrix}
        0 & r_{\rm A}^* \\ r_{\rm A} & 0
    \end{pmatrix}
\end{equation}
where $\alpha (\epsilon ) = \sqrt{1-\epsilon^2/\Delta_0^2} + i \epsilon / \Delta_0$, and
\begin{equation}\label{eq:RA_AndreevApproximation}
    r_{\rm A} = \alpha(\epsilon) \begin{pmatrix}
        i e^{i \phi/2} & 0 \\ 0 & i e^{-i \phi/2}
    \end{pmatrix},
\end{equation}
with $|\Delta| = \Delta_0$ being the modulus of the superconducting order parameter and $\phi$ denoting the superconducting phase difference between the two superconducting regions.
In Eq.~\eqref{eq:RA_AndreevApproximation}, the normal reflection at the interface can be neglected in the Andreev approximation limit~ \cite{Beenakker_1991_a}, 
where the Fermi level of each superconducting sector is much larger than the superconducting gap parameter $\Delta_0$.
In the central scattering region, electron and hole degrees of freedom are decoupled and, in the same basis choice as before, $s_{\rm N}(\epsilon)$ is expressed in block diagonal form for the two subspaces as \cite{Beenakker_1991_a,vanHeck_2014_a}
\begin{equation}\label{eq:SN_NoApproximation}
    s_{\rm N}(\epsilon) = \begin{pmatrix}
        s(\epsilon) & 0 \\ 0 & s(-\epsilon)^*
    \end{pmatrix}.
\end{equation}
In the short junction regime, where the coherence length is much longer than the length of the junction, $\xi \sim \hbar v /\Delta_0 \gg L$, and the Thouless energy, $\hbar v/L$, becomes the dominant energy scale of the system, we can further approximate $s(\epsilon) \approx s(-\epsilon) \approx s(0)$ in Eq.~\eqref{eq:SN_NoApproximation} \cite{Beenakker_1991_a}.
Finally, after some algebraic manipulations, the \gls*{abs} energy, $\epsilon$, can be found by solving the following eigenvalue problem \cite{vanHeck_2014_a,Irfan_2018_a,Chirolli_2025_a}
\begin{subequations}\label{eq:ABS_Eigenproblem}
\begin{align}
    A^+ A \Psi_{\rm in}^e &= \frac{\epsilon^2}{\Delta_0^2} \Psi_{\rm in}^e, \label{eq:ABS_Eigenproblem_a} \\
    A &= \frac{1}{2} \left( r_{\rm A} s(0) - s^{\rm T}(0) r_{\rm A} \right)~, \label{eq:ABS_Eigenproblem_b}
\end{align}
\end{subequations}
from which the supercurrent flowing through the junction can be computed as
\begin{align}\label{eq:Supercurrent_FromA}
    I (\phi) & = -\frac{e^2}{\hbar} \sum_p \frac{d \epsilon_p}{d \phi}, \\
    \frac{d \epsilon_p}{d \phi} & = \frac{\Delta_0^2}{2\epsilon} \left\langle \Psi_{\rm in}^e \left| \frac{d \left( A^T A \right)}{d \phi} \right| \Psi_{\rm in}^e \right\rangle,\end{align}
where the contribution of the continuum spectrum, for $|E| > \Delta_0$, can be neglected in the short junction regime \cite{Beenakker_1991_a}.
If the scattering region of the junction is time-reversal symmetric, the problem can be further simplified to find \cite{Beenakker_1992_a}
\begin{gather}
    \epsilon_p = \Delta_0 \sqrt{1-\tau_p \sin^2(\phi /2)} \label{eq:ABS_BeenakkerFinal}~, \\
    I(\phi) = \frac{2e \Delta_0}{\hbar} \sum_{p} \frac{\tau_p \left(k \right) \sin \phi}{4 \sqrt{1-\tau_{p} \left(k \right) \sin^2 (\phi/2)}}~, \label{eq:Supercurrent_BeenakkerFinal}
\end{gather}
where $\tau_p$ are the eigenvalues of the Hermitian matrix $t_{\rm LR}t_{\rm LR}^\dagger$, where $t_{\rm LR}$ is the transmission matrix from the right to the left lead, which is obtained from $s(0)$
\begin{equation}
    s(0) = \begin{pmatrix}
        r_{\rm LL} & t_{\rm LR} \\ t_{\rm RL} & r_{\rm RR}
    \end{pmatrix}.
\end{equation}
Eqs.~(\ref{eq:ABS_BeenakkerFinal}-\ref{eq:Supercurrent_BeenakkerFinal}) directly connect the transmission probabilities of a normal graphene junction to the supercurrent in a \gls*{gjj}.

The tight-binding Hamiltonian of the inner graphene layer, which includes the SOC terms induced by the substrate, is expressed as~\cite{Konschuh_2010_a,Kochan_2017_a,Frank_2018_a}
\begin{equation} \label{eq:Hamiltonian}
\begin{aligned}
    \mathcal{H} & = - t \sum_{\langle i,j\rangle,s} c_{is}^\dagger c_{js} - \mu \sum_{i,s} c_{is}^\dagger c_{is} \\ 
    & + \frac{2 i \lambda_{\rm R}}{3} \sum_{\langle i,j \rangle,s,s'} \left[ \left( \hat{\bm{s}} \times \bm{d}_{ij} \right)_z \right]_{ss'} c_{is}^\dagger c_{js'} \\
    & + \frac{i}{3\sqrt{3}} \sum_{\langle\langle i,j \rangle\rangle,s,s'} \lambda_{\rm I}^i \nu_{ij} \left[\hat{\bm{s}}_z\right]_{ss'} c_{is}^\dagger c_{js'},
\end{aligned}
\end{equation}
where $c^\dagger_{is}$ ($c_{is}$) creates (annihilates) an electron on site $i$ with spin z-projection $s$, $\hat{\bm{s}}$ is the Pauli operator describing the spin degree of freedom, $\bm{d}_{ij}$ is the real space vector connecting site $j$ to site $i$, and $\nu_{ij} = \pm$ based on whether in hopping from site $j$ to site $i$ an electron follows a clockwise or counter-clockwise path along a hexagonal ring; $\langle i,j \rangle$ indicates the sum over nearest neighbors and $\langle\langle i,j \rangle\rangle$ that over next-nearest neighbors.
In addition, in Eq.~\eqref{eq:Hamiltonian}, $t$ denotes the nearest-neighbor hopping amplitude in graphene and $\mu$ represents the Fermi energy; $\lambda_{\rm R}$ characterizes the magnitude of the \gls*{rb}, while $\lambda_{\rm I}^{A,B}$ specifies the strength of the intrinsic SOC, which may take different values on the two sublattices~\cite{Frank_2018_a}.
In the following, we will consider the combinations $\lambda_{\rm KM} = (\lambda_{\rm I}^A + \lambda_{\rm I}^B)/2$, known as \gls*{km}, and $\lambda_{\rm VZ} = (\lambda_{\rm I}^A - \lambda_{\rm I}^B)/2$, known as \gls*{vz}.
A dominant \gls*{km} drives the layer in the famous \gls*{qsh} phase of graphene.
While a combination of $\lambda_{\rm VZ}$ and $\lambda_{\rm R}$ \glspl*{soc} leads the graphene layer to the newly proposed \gls*{qvsh} phase \cite{Frank_2018_a}, which will be the main focus of this work.

Operatively, we implement the Hamiltonian \eqref{eq:Hamiltonian} and compute the scattering matrix, $s(0)$, of the graphene layer using the {\sc kwant} Python package \cite{Groth_2014_a}.
In the later parts of this work, we will numerically generate a double junction of width $W = 295a$, where $a$ is the lattice constant of the honeycomb lattice, and a scattering region with a length of $L = W/5$.
Our goal is to simulate setups of comparable ratios and lengths of around $200$ nm \cite{Wakamura_2020_a}, which are within the short junction approximation while still behaving ballistically, and for this reason, we consider all energies in units of Thouless energy $\hbar v /L$.

More details on the numerical calculations can be found in the Appendix \ref{sec:App:simulations_details}.

\section{Normal state single junction}\label{sec:NormalState_Transport}

\begin{figure*}[t!]
\centering
\raisebox{2em}{
\begin{overpic}[width=0.35\linewidth]{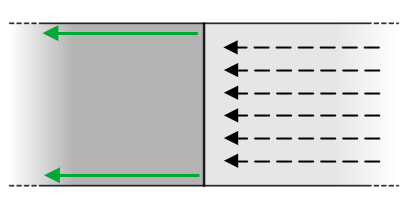}\put(-7,55){a)}\end{overpic}}\hspace{1em}
\begin{overpic}[width=0.5\linewidth]{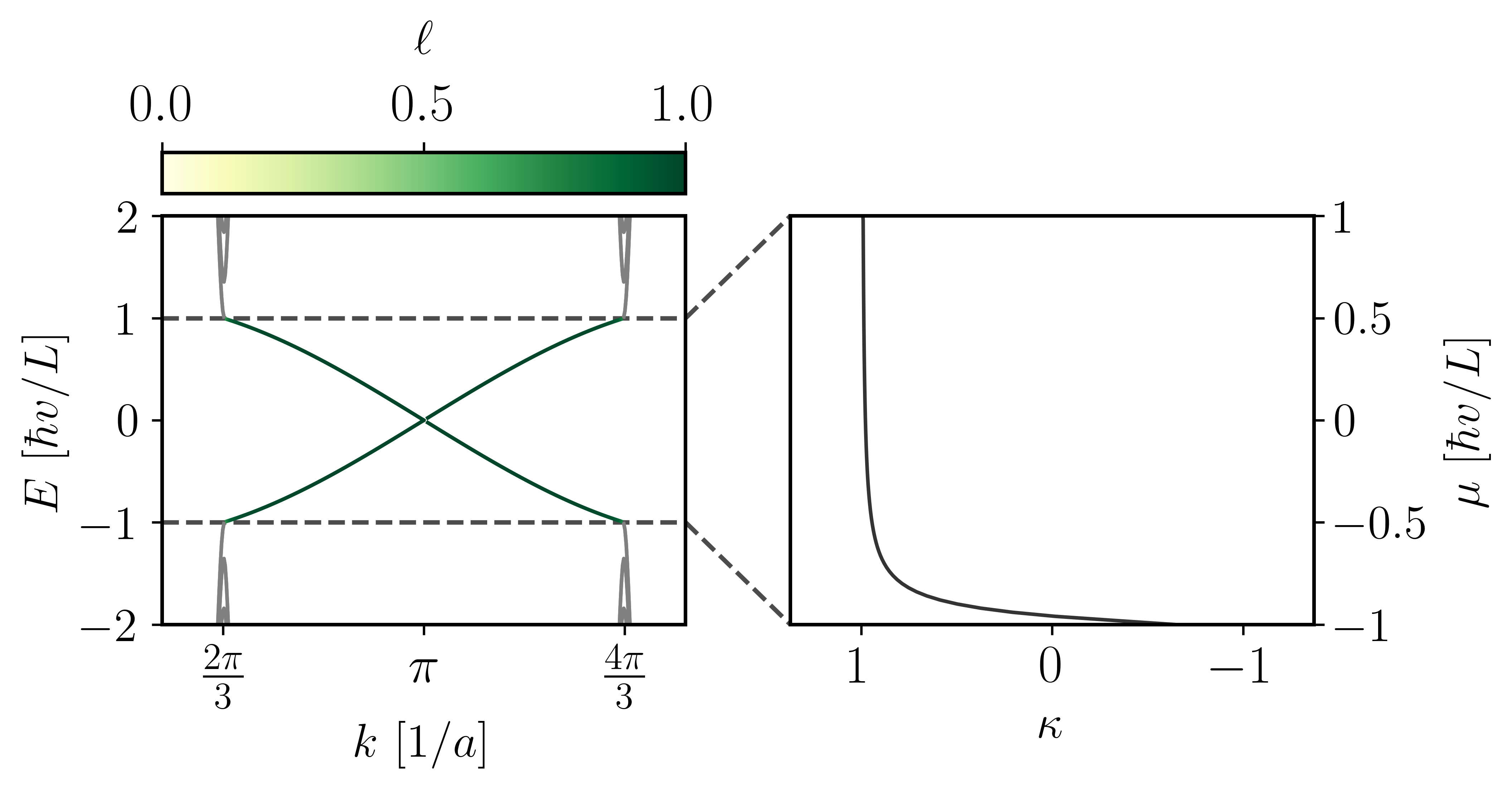}\put(0,46){b)}\end{overpic}
\caption{a) schematic of a single junction where the left sector is made of zigzag graphene ribbon in the \gls*{qsh} phase with $\lambda_{\rm KM} = \hbar v /L$, and the right sector is made of pristine graphene with Fermi level $\mu_R = 0.2t$; in green we highlighted the left-moving helical edge states, while the current is injected from the right sector and pictorially represented by black dashed arrows. The width of the junction is $W=295a$. 
b) the left panel shows the electronic band structure of a \gls*{qsh} graphene ribbon, where the scale of greens represents the degree of localization of the helical edge states, as defined in Eqs.~\eqref{eq:Localization_Definition}-\eqref{eq:Localization_Definition_Weighting}, and the dashed black horizontal lines highlight the energy gap; the energies are measured in units of $\hbar v /L$ and the momenta in units of $1/a$.
The right panel shows the valley polarization of the incoming current, $\kappa$, as defined in Eq.~\eqref{eq:IncomingValleyPolarization}, for Fermi level values, $\mu$ (in units of $\hbar v /L$), inside the energy band gap. The $\kappa$ axis is reversed so as to align with the appearance of the $K = 2\pi/3~[1/a]$ and $K'= 4\pi/3~[1/a]$ valleys in the left panel.}
\label{fig:JP_VP_KM}
\end{figure*}

\begin{figure*}[t!]
\centering
\raisebox{2em}{
\begin{overpic}[width=0.35\linewidth]{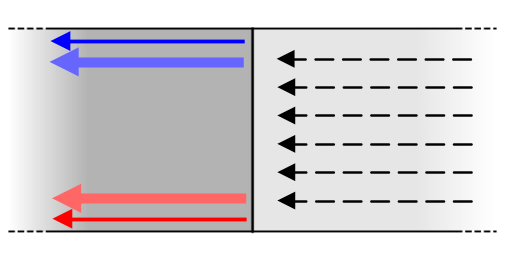}\put(-7,55){a)}\end{overpic}}\hspace{1em}
\begin{overpic}[width=0.52\linewidth]{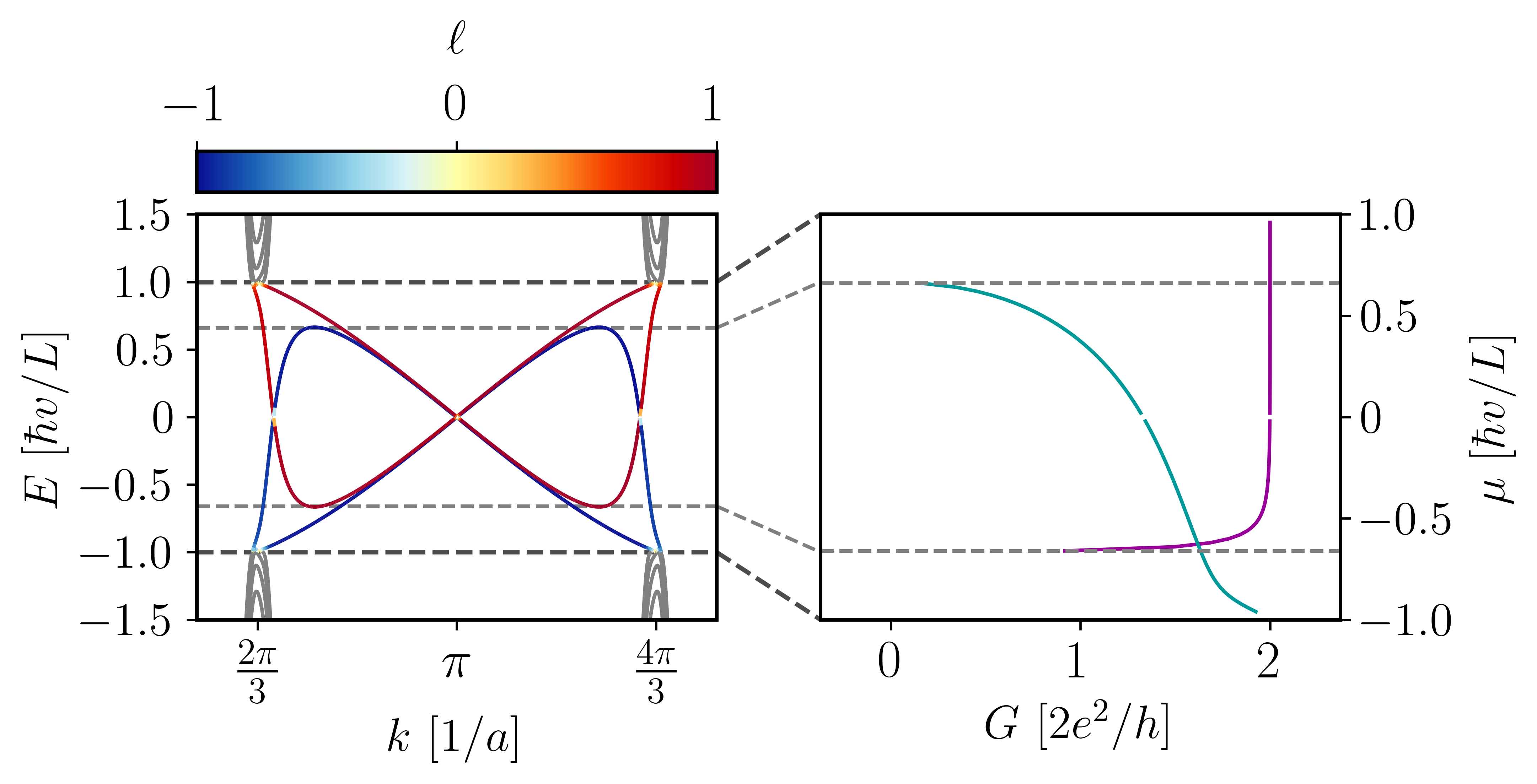}\put(0,44){b)}\end{overpic}
\caption{a) schematic of a single junction where the left sector is made of a zigzag graphene ribbon in the \gls*{qvsh} phase with $\lambda_{\rm VZ} = \hbar v /L$ and $\lambda_{\rm R} = 5~ \hbar v/L$, in red (blue) the electron modes of the bottom (top) edge, with solid [pale] color indicating the pseudohelical [valley] edge states.
The right sector is made of pristine graphene with Fermi level $\mu_R = 0.2t$, and the current is injected from the right side and pictorially represented by black dashed arrows. The width of the junction is $W=295a$. 
%
In b) the left panel shows the band structure of the \gls*{qvsh} phase graphene ribbon: the color scale indicates the degree of edge localization, defined in Eqs.~\eqref{eq:Localization_Definition} and \eqref{eq:Localization_Definition_WeightingDifferent}, in red (blue) for the bottom (top) edge, the dashed black horizontal lines highlight the energy band gap, and the dashed gray horizontal lines denote the energy range in which the edge states exist on both edges of the junction; the energies are measured in units of $\hbar v /L$ and the momenta in units of $1/a$. The right panel shows the edge-resolved conductance, calculated using Eq.~\eqref{eq:EdgeResolvedConductance}, for Fermi levels values, $\mu$ (in units of $\hbar v /L$), inside the energy gap, in magenta for the bottom edge, $G^{\rm B}$, and in teal for the top one, $G^{\rm T}$.}
\label{fig:JP_EC_VZ}
\end{figure*}

As initial step in the analysis of the superconducting transport of the \gls*{gjj} under study, we first consider a single junction with an interface between non-superconducting graphene ribbons in different phases.
Specifically, the right sector is composed of pristine graphene ($\lambda_{\rm R} = \lambda_{\rm KM} = \lambda_{\rm VZ} = 0$) with a high Fermi level of $\mu_{\rm R} = 0.2 t$, where $t$ is the hopping parameter appearing in Hamiltonian \eqref{eq:Hamiltonian}.
For the left sector, we consider two cases: graphene in the \gls*{qsh} phase, in Fig.~\ref{fig:JP_VP_KM}~a), and graphene in the \gls*{qvsh} phase, in Fig.~\ref{fig:JP_EC_VZ}~a).
The transport properties are computed for current injected from the right sector into the left one and are limited to zigzag-terminated ribbons.

\subsection{QSH Phase - Valley Filtering Effect}\label{sec:QSH_SingleJunction_ValleyFiltering}

A graphene ribbon in the \gls*{qsh} phase is characterized by localized edge states whose energies reside within the energy band gap \cite{Kane_2005_a,Kane_2005_b}.
These edge states are spin-polarized, with different electron spins moving in opposite directions on the same edge and in the same direction on opposite edges.
For this reason, they are known as \textit{helical} edge states, emphasizing the connection between the electron spin and the direction of propagation.
Fig.~\ref{fig:JP_VP_KM}~a) illustrates a schematic of the junction where only the left-moving states are shown, in green for the edge states of the \gls*{qsh} phase and in dashed black for the bulk ones of pristine graphene.
The left panel of Fig.~\ref{fig:JP_VP_KM}~b) shows the electronic band structure of a zigzag graphene ribbon of width $W=295a$ in the \gls*{qsh} phase, with $\lambda_{\rm KM} = \hbar v/L$, which also corresponds to the energy band gap.
The scale of greens represents the degree of localization of the helical edge states that we calculated as~\cite{Frank_2018_a}
%
\begin{align}
    \ell(E,k) & = \sum_{i} l(y_i) \left| \psi_i (E,k) \right|^2, \label{eq:Localization_Definition} \\
    l(y_i) &= \begin{cases}
        - \dfrac{2 y_i}{W} & \text{for}\quad -\dfrac{W}{2} \leq y_i \leq 0 \\
        \dfrac{2 y_i}{W} & \text{for} \quad 0 \leq y_i \leq \dfrac{W}{2}
    \end{cases}, \label{eq:Localization_Definition_Weighting}
\end{align}
where $l(y_i)$ is a positive weighting function that attributes a higher weight to local densities of the wavefunction closer to the edges.

Here, we focus on the valley degree of freedom when analyzing the transport properties across the junction.
Specifically, we study the valley polarization of the incoming current for the electronic states that solve the transport problem.
Mathematically, the valley polarization of the incoming current is calculated in the following way.
By numerically solving the continuity problem along the interface, we obtain the transmission matrix $t_{\rm LR}(\mu)$ from the right to the left sector, for a given Fermi level of the \gls*{qsh} phase graphene, $\mu$.
From $t_{\rm LR}\left( \mu \right)$ we can then compute the transmission probabilities, $\tau\left( \mu \right)$, across the junction by solving the eigenvalue problem \cite{Beenakker_1991_a}
%
\begin{equation}\label{eq:TransmisssionEigenproblem_Original}
    \left[ t_{\rm LR} \left( \mu \right) ^\dagger t_{\rm LR} \left( \mu \right) \right] \psi_\tau = \tau \left( \mu \right) \psi_\tau ,
\end{equation}
where $\psi_\tau = \left[ \psi_\tau^{k_1s_1}, \cdots, \psi_\tau^{k_i s_i}, \cdots\right]^T$ is a vector with the components of the total incident wavefunction relative to basis of incoming propagating modes from the pristine sector, which can be characterized by their crystal momentum, $k_i$, and their spin z-projection, $s_i$.
Finally, we can define the valley polarization of the incoming current as
\begin{equation}\label{eq:IncomingValleyPolarization}
    \kappa \left( \mu \right) = \frac{1}{\sum_\tau \tau} \sum_{\tau,i} \tau |\psi^{k_i s_i}_\tau| ^2 
\,{\rm sgn}[(K+K')/2-k_i],
\end{equation}
where the sum extends to all transmission eigenvalues at the Fermi level $\mu$, $K = 2\pi /3 \left[ 1/a \right]$ and $K' = 4\pi /3 \left[ 1/a \right]$, and the sign function is used to effectively label an incoming propagating mode as belonging to a given valley using its vicinity to it in momentum space.
Here,  $\left| \kappa\left( \mu \right) \right|$ represents the intensity of valley polarization, $0$ ($1$) being the completely unpolarized (polarized) case, and ${\rm sgn}\left( \kappa \left( \mu \right)\right)$ tells us which valley is preferred, positive [negative] for the $K$ [$K'$] valley.
%
%
%

The right panel of Fig.~\ref{fig:JP_VP_KM}~b) shows the valley polarization of the incoming current as a function of the Fermi level, $\mu$, for a single junction of width $W=295a$ between graphene in the \gls*{qsh} phase, with $\lambda_{\rm KM} = \hbar v/L$, and pristine graphene, with $\mu_R = 0.2t$, both terminating in a zigzag edge.
We see that for most Fermi level values inside the energy band gap, the incoming current gets completely valley polarized by the interface between the two different graphene phases.
This means that electrons predominantly from the $K$ valley are transmitted through the helical edge states of the \gls*{qsh} ribbon.
Moreover, we find transmission eigenvalues that are consistently over $\tau \left( \mu \right) \gtrsim 0.995$ inside the energy gap, as expected due to the impossibility of backscattering, a characteristic of the \gls*{qsh} helical edge states \cite{Kane_2005_a,Kane_2005_b}.
Together, these mean that a long enough \gls*{qsh} graphene stripe, so as to avoid evanescent tunneling, effectively acts as a valley-filtering device between two pristine zigzag graphene ribbons; more details on the double junction setup can be found in Appendix~\ref{sec:App:QSH_DoubleJunction}.
This resembles the valley-filtering device, based on a restriction in a zigzag graphene ribbon, proposed in Ref.~\cite{Rycerz_2007_a}.

\subsection{QVSH Phase - Asymmetric Edge Transport}\label{sec:QVSH_AsymmetricTransport}

Combining \gls*{vz} and \gls*{rb}, the graphene ribbon is driven into a topologically trivial gapped phase that is still characterized by the existence of dispersive localized edge states, known as the \gls*{qvsh} phase \cite{Frank_2018_a}.
The edge behavior of a ribbon in the \gls*{qvsh} phase is even richer than that in the \gls*{qsh} phase but is restricted to those edge terminations that do not mix the degrees of freedom of the graphene valleys.
Specifically, a zigzag graphene ribbon in the \gls*{qvsh} phase hosts two pairs of edge states per edge \cite{Frank_2018_a}.
One pair is composed of extremely localized \textit{pseudohelical} edge states: they move in opposite directions for different electron spin z-projections on the same edge but in the same direction for opposite edges (in contrast to the \gls*{qsh} case).
The other pair of edge states is spin-unpolarized and much less localized on the edges. Moreover, due to their momentum space localization close to the graphene valleys, they were dubbed \textit{valley-edge modes}.
Fig.~\ref{fig:JP_EC_VZ}~b) shows the electronic band structure of a zigzag graphene ribbon of width $W=295a$ with $\lambda_{\rm VZ} = \hbar v/L$ and $\lambda_{\rm R} = 5 ~ \hbar v/L$; more details on the \gls*{soc} values can be found in Appendix~\ref{sec:App:simulations_details}.
The dashed horizontal black lines confine the bulk energy band gap, given by $2\min\left( \lambda_{\rm VZ},\lambda_{\rm R} \right)$, and the color scale from blue to red shows the degree of edge localization of the intragap states, which is computed as in Eq.~\eqref{eq:Localization_Definition} 
but using a different weighting function
%
\begin{equation}\label{eq:Localization_Definition_WeightingDifferent}
    l(y_i) = - \dfrac{2 y_i}{W},
\end{equation}
which, in contrast to Eq.~\eqref{eq:Localization_Definition_Weighting}, now provides both positive and negative weights, allowing us to distinguish between the bottom (positive weight) and the top (negative weight) edges.
Fig.~\ref{fig:JP_EC_VZ}~a) shows a schematic of the left-moving modes in the single junction. Here, red (blue) arrows denote the electron modes of the bottom (top) edge, and solid [pale] colors indicate the pseudohelical [valley] edge states.

We now compute the edge-resolved conductance for a single junction, composed of zigzag graphene ribbons of width $W=295a$, where the left sector is in the \gls*{qvsh} phase, with $\lambda_{\rm VZ} = \hbar v/L$ and $\lambda_{\rm R} = 5~\hbar v/L$, and the right sector in the pristine phase, with $\mu_R = 0.2t$.
We compute the edge-resolved conductance in the following way.
Similarly to the previous section, we solve the transposed eigenvalue problem to that of Eq.~\eqref{eq:TransmisssionEigenproblem_Original}
\begin{equation}\label{GJJE:eq:TransmisssionEigenproblem_Transposed}
    \left[ t_{\rm LR} \left( \mu \right) t_{\rm LR} \left( \mu \right) ^\dagger \right] \psi_\tau = \tau \left( \mu \right) \psi_\tau
\end{equation}
%
%
%
where the components of $\psi_\tau$ are now relative to the basis of the outgoing modes in the \gls*{qvsh} sector, which consist only of edge states (for $|\mu| \leq |\lambda_{\rm VZ}|$).
We can then collectively label as $\psi_\tau^{\rm T}$ ($\psi_\tau^{\rm B}$) the components of $\psi_\tau$ if they refer to edge states localized on the top (bottom) edge, with $\ell < 0$ ($\ell > 0$).
After solving the eigenvalue problem, we finally define the edge-resolved conductance as
\begin{equation}\label{eq:EdgeResolvedConductance}
    G = \begin{pmatrix}
        G^{\rm T} \\ G^{\rm B}
    \end{pmatrix} = \frac{2 e^2}{h} \sum_\tau \tau \begin{pmatrix}
        |  \psi^{\rm T}_\tau |^2 \\ |  \psi^{\rm B}_\tau |^2
    \end{pmatrix}.
\end{equation}
The results are shown in the right panel of Fig.~\ref{fig:JP_EC_VZ}~b), in magenta for the bottom edge, $G^{\rm T}$, and in teal for the top edge, $G^{\rm B}$.
A level of asymmetry between the two edges is expected in these systems because the \gls*{soc} interaction breaks the mirror symmetry of pristine zigzag graphene in the transverse ribbon direction.
For example, in the energy regions confined between the black and gray dashed horizontal lines, the \gls*{qvsh} ribbon itself hosts propagating modes only on one edge of the ribbon, on the bottom (top) edge for the positive (negative) energy range.
%
Based on the electronic band structure shown in Fig.~\ref{fig:JP_EC_VZ}~b), the localization properties of the edge states within the gap are mapped onto each other under the transformation $\mu \to -\mu$. Therefore, one might expect the transport behavior observed at the Fermi level $\mu$ to be mirrored, with the asymmetry reversed, at the opposite Fermi level $-\mu$.
However, this is not what our numerical results show.
In the right panel of Fig.~\ref{fig:JP_EC_VZ}~b), the transport is highly asymmetric; in particular, across nearly the entire energy-gap range, one edge (here, the bottom edge) is much more strongly favored than the other.
The explanation for this behavior is the following.
%
In the previous section, we showed that the helical edge states of a \gls*{qsh} ribbon almost entirely polarize the injected current into a single valley. Interestingly, the quasihelical edge states of a \gls*{qvsh} ribbon lead to a very similar form of valley-selective transport.
In turn, the valley-edge state of the preferred valley has a better matching condition compared to that of the other valley. 
In this way, the \gls*{qvsh} ribbon, due to the presence of valley-localized edge states, converts the valley polarization of the incoming current into edge polarization, leading to asymmetric edge transport.
%
Moreover, we note that reversing the external doping of the pristine sector, as $\mu_{\rm R} \to - \mu_{\rm R}$, will change the valley polarization of the helical and pseudohelical edge states as $\kappa(\mu) \to -\kappa(-\mu)$ and, as a consequence, also the edge polarization in the transport problem just analyzed.
See Appendix \ref{sec:App:QVSH_ValleyPolarization} for a further analysis of the valley polarization of the incoming current in this setup.

\section{Edge Carried Supercurrent in a GJJ}\label{sec:Edge_Supercurrent}

\begin{figure*}[t!]
\centering

\begin{overpic}[width=0.3\textwidth,trim={0 0cm 0 0cm}]{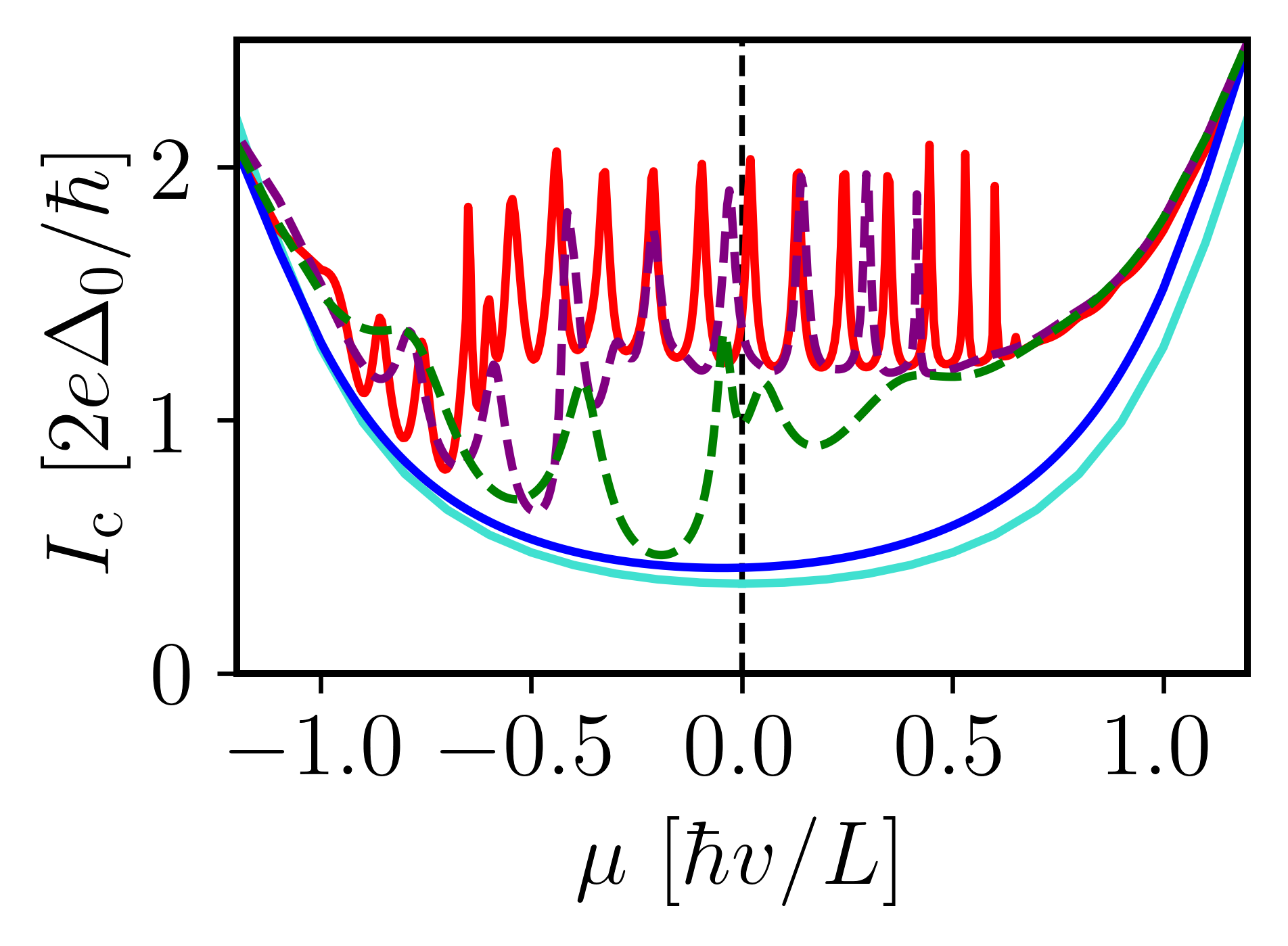}\put(2,70){a)}\end{overpic}
\begin{overpic}[width=0.3\linewidth]{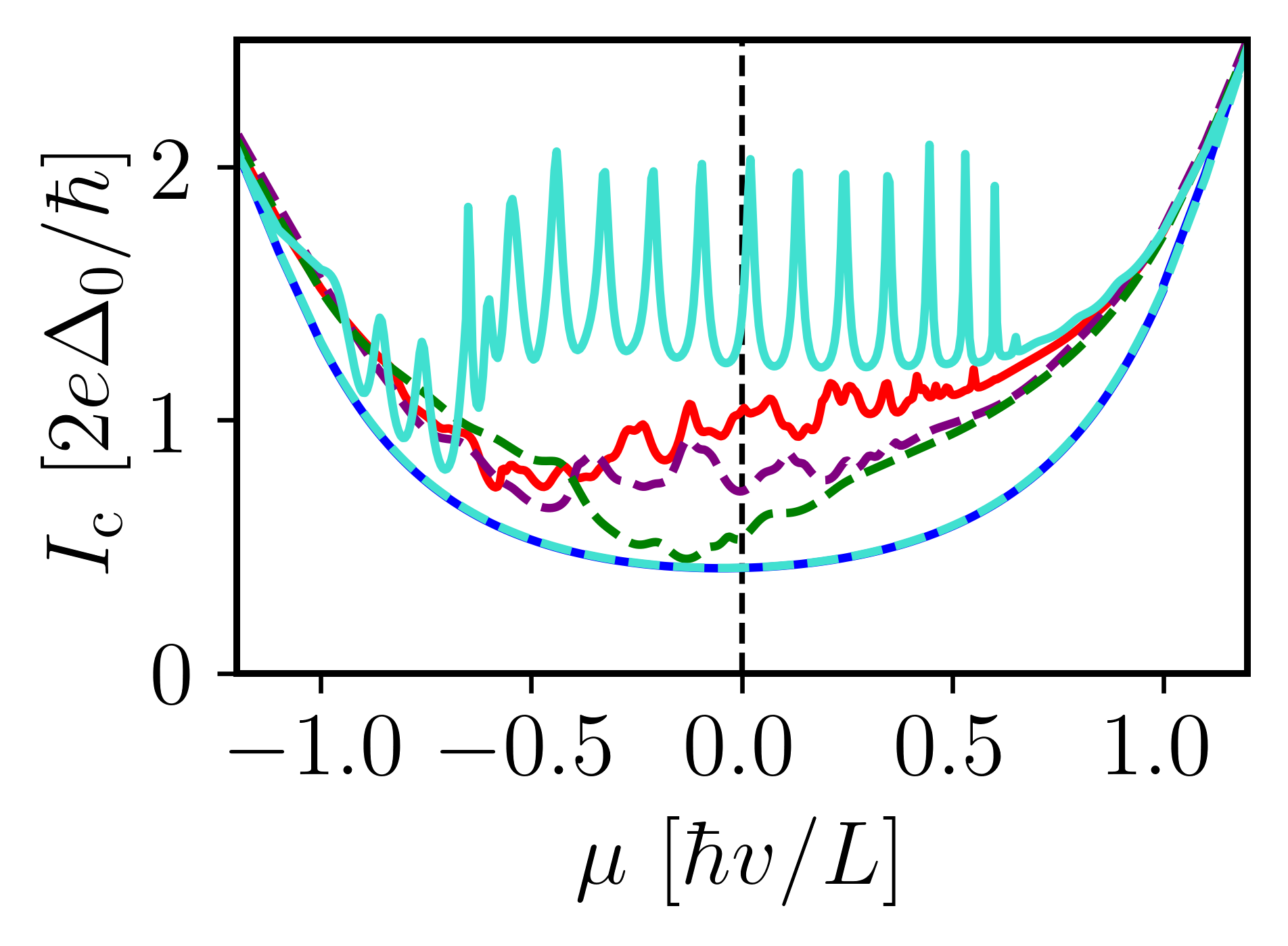}\put(2,70){b)}\end{overpic}
\begin{overpic}[width=0.3\textwidth,trim={0 0cm 0 0cm}]           {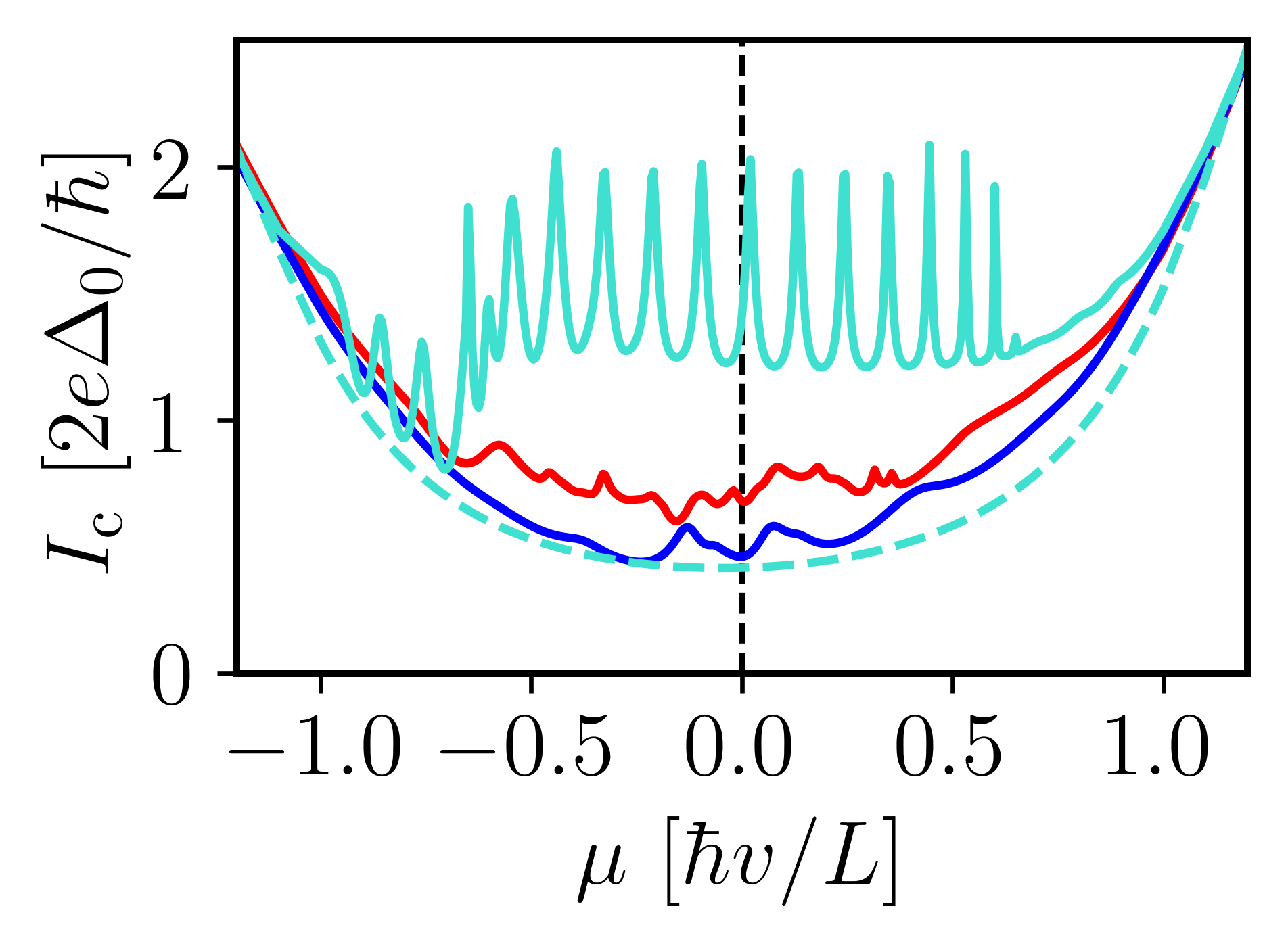}\put(2,70){c)}\end{overpic}

\caption{a) critical current $I_{\rm c}$ (in units of $2 e \Delta_0/\hbar$) of the \gls*{gjj} as a function of the Fermi level $\mu$ (in units of $\hbar v/L$) of the \gls*{qvsh} graphene region; the parameters used are $\lambda_{\rm R} = 5~ \hbar v/L$, $\lambda_{\rm VZ} = \hbar v /L$ and a junction length and width of $L=59a$ and $W=5L$. In red the results for a pure zigzag ribbon with $(m,n)=(1,0)$, in purple (green) those for a $(m,n)=(2,1)$ ($(m,n)=(1,2)$) termination and in blue the ones for a pure armchair ribbon, $(m,n)=(0,1)$. As a reference, we show in cyan the results of a previous work where the critical current is computed analytically using periodic boundary conditions \cite{Bonasera_2025_a}, effectively corresponding to the limit of an infinite $W$.
b) same as in a) but for light disordered edges, as described in the main text; the color coding is the same as in a), except for the cyan lines that, as reference, show the critical current computed in a) for the zigzag (solid line) and armchair (dashed line) cases.
c) same as in b) but for heavy disordered edges, as described in the main text.}
\label{fig:Crit_Current_R5S1_CleanAndDisordered}
\end{figure*}
%


Building on the previous results, we now move to the study of superconducting transport in a \gls*{gjj} made of two superconducting graphene leads separated by a scattering layer of \gls*{qvsh} graphene.
In particular, we focus on the effects of the edge states on the supercurrent.
We characterize the current phase relation, $I \left(\phi\right)$, of the junction by its maximum value, known as the critical current
\begin{equation}\label{eq:CriticalCurrent_Definition}
I_{\rm c} = \max_{\phi} \left| I \left(\phi \right) \right|,
\end{equation}
which represents the maximum current that the junction can sustain without developing a voltage difference across the leads \cite{Tinkham_1996_a}.
%

We consider a junction of the same width and \gls*{soc} parameter strengths as in the previous section with a length of $L=W/5$.
In Fig.~\ref{fig:Crit_Current_R5S1_CleanAndDisordered}~a) we show the results for the critical current against the Fermi level of the inner graphene layer, $I_{\rm c}\left(\mu\right)$, for different kinds of edge termination.
We limit ourselves to a Fermi level range close to the energy gap to focus on the edge contribution and we characterize an edge termination based on its zigzag ($m$) and armchair ($n$) content, defined by the couple $(m,n)$; more details on the nomenclature for the different graphene terminations can be found in Appendix \ref{sec:App:simulations_details}.
Specifically, we plot in red the results for a pure zigzag edge, $(m,n)=(1,0)$, in purple those for a mostly zigzag termination, $(m,n)=(2,1)$, in green those for a mostly armchair one, $(m,n)=(1,2)$, and in blue those of a pure armchair edge, $(m,n)=(0,1)$; as a reference, we plot in cyan the results of a previous work where the current is computed analytically
using periodic boundary conditions \cite{Bonasera_2025_a}, where the terminations are irrelevant.
From Fig.~\ref{fig:Crit_Current_R5S1_CleanAndDisordered}~a), we can make a few observations.
First, we see that, as expected, the critical current for the proper armchair termination closely follows the one we computed analytically in the previous work, for a value of $W/L=5$, originating only from bulk states.
For non-pure armchair terminations, instead, the contribution of the edge states is dominant within the energy gap.
Moreover, we can divide this contribution into two parts.
We know that every transmission channel with unit transmission probability, $\tau\approx 1$, contributes to the supercurrent a factor of $\frac{e\Delta_0}{\hbar}$ \cite{Beenakker_1991_a}.
One part of the edge contribution then comes from two channels with almost unit transmission probability:
these belong to the higher transmission edge of the junction (the bottom, red one in Fig.~\ref{fig:JP_EC_VZ}) via hybridization of the pseudohelical and valley edge states.
The same hybridization also occurs on the lower-transmission edge and, due to the coherent nature of the scattering problem, produces two transmission channels that undergo cycles of constructive interference with varying chemical potential.
This accounts for the second part of the edge contribution.
The subdivision between high and low transmission edge channels is very evident when looking at the zigzag result for $(m,n)=(1,0)$, shown in red:
there is a wide region of chemical potential values, for $|\mu| \lesssim 0.7 \hbar v/L$, where we have both contributions, while for $0.7 \hbar v/L \lesssim \mu \lesssim \hbar v/L$ ($-\hbar v/L \lesssim \mu \lesssim - 0.7\hbar v/L$) only the one from the high (low) transmission edge exists.
This behavior aligns perfectly with the description of the edge states in a \gls*{qvsh} ribbon of Fig.~\ref{fig:JP_EC_VZ}~b): for $ 0.7\hbar v/L \lesssim \mu \lesssim \hbar v/L$ the scattering region only contains the high transmissive edge states of the bottom edge (in red), while for $ -\hbar v/L \lesssim \mu \lesssim -0.7\hbar v/L$ it contains only the low transmissive ones of the top edge (in blue).

For the edge terminations with increased armchair content, we find a similar effect with two main differences: the tunnel constructive interference peaks become sparser, and the energy window where both contributions coexist becomes smaller.
In general, the condition of constructive interference is met when the wavefunction accumulates a phase multiple of $2\pi$ in a round-trip across the junction.
Neglecting the phase accumulation at the interfaces, this means that
$ k L = n \pi$, where $k$ is the channel crystal momentum and $n$ is an integer.
In our case we have $k \in [\Lambda_{<}^ {(m,n)},\Lambda_{>}^ {(m,n)}]$, which defines the momentum range where the edge states exist, leading to interference peaks for $n=1,\ldots, n^{(m,n)}$ with $n^{(m,n)} \sim L \left| \Lambda^{(m,n)}_>-\Lambda^{(m,n)}_< \right|/(2\pi) \equiv L \Lambda^ {(m,n)}/(2\pi)$, which gives an estimate of the total number of transmission peaks for a given edge termination $(m,n)$.
In the energy domain, the periods of the interference cycle are determined by the group velocities of the edge states.
%
With increasing armchair content, the edge states acquire a steeper energy dispersion and, because their energies must remain within the bulk gap, the corresponding momentum window shrinks, reducing the number of allowed interference resonances and thus the number of transmission peaks observed numerically.
More details on this can be found in Appendix \ref{sec:App:EnergyDispersion_MoreArmchair}.

\subsection{Effects of Edge Disorder}

Here, we analyze the robustness of the edge contribution against disorder in the form of edge roughness.
We consider the case of small disorder for the $(m,n)=\{(1,0),(2,1),(1,2),(0,1)\}$ terminations and of heavy disorder for the $(m,n)=\{(1,0),(0,1)\}$ ones.
For the former, when building the tight-binding graphene layer, we introduce a probability of $20 \%$ that an edge atom, defined as having fewer than three nearest-neighbor atoms, is removed.
After this process, the edge is cleaned of dangling atoms, defined as having fewer than two nearest-neighbor hopping atoms \cite{Groth_2014_a}, also known as Klein defects \cite{Ivanovskaya_2012_a}.
For the latter, we repeat the previous process six times, which, on average, yields a maximum damage depth of $3$-$4$ $a$ at the edge, slightly more than $1\%$ of the junction width in our calculations.
We then calculate the $I_{\rm c}(\mu)$ plots for ten different realizations of disorder for every edge termination considered and average the results.

The computed averages are shown in Fig.~\ref{fig:Crit_Current_R5S1_CleanAndDisordered}~b) for the small disorder cases and in Fig.~\ref{fig:Crit_Current_R5S1_CleanAndDisordered}~c) for the heavy disorder ones.
In both figures, we include as a reference the clean zigzag edge result in solid cyan and the clean armchair edge one in dashed cyan.
In Fig.~\ref{fig:Crit_Current_R5S1_CleanAndDisordered}~b), we see that, for edge terminations with a zigzag content, the disorder reduces the edge state contribution to the critical current.
Specifically, the contribution from the highly transmissive edge is only slightly reduced, but the one from the low-transmission one is drastically lowered because most of the current peaks, raised because of the constructive interference, are smoothed out by the averaging procedure.
For each specific realization of disorder, the critical current continues to exhibit some constructive interference, although its structure is less pronounced than in the clean case; further details are provided in Appendix \ref{sec:App:CriticalCurrent_DisorderedConfigurations}.
In contrast, armchair termination appears to be completely insensitive to small disorder.

The situation changes when we consider the heavy disorder.
In this situation, the difference in edge contributions to the supercurrent between disordered zigzag (red) and armchair (blue) terminations becomes smaller, as shown in Fig.~\ref{fig:Crit_Current_R5S1_CleanAndDisordered}~c), and the armchair edge itself begins to support a nonzero contribution of edge-state transport; this represents a case of disorder-induced transport.
Indeed, as we saw previously, a clean armchair sample normally does not host edge modes in the \gls*{qvsh} phase of graphene, and all the transport occurs through evanescent bulk modes.
Disorder creates zigzag-edge defects within an armchair termination, leading to localization of the wavefunction and enhancing its edge transport properties.
This is a similar result to what was previously obtained in gapped bilayer graphene, 
where the authors found a universal value of the subgap conductance for strong enough edge disorder, independently of the starting edge configuration \cite{Li_2011_a}.

Overall, we found that even though the edge contribution in a \gls*{qvsh} graphene ribbon is not topologically protected, it is still fairly robust against non-magnetic disorder.
Moreover, when the disorder is strong enough, it can actually enhance the edge transport in previously non-conducting armchair terminations.

\section{Magnetic interference pattern}\label{sec:MIP}

\begin{figure*}[t!]
\centering

\begin{overpic}[width=0.35\linewidth]{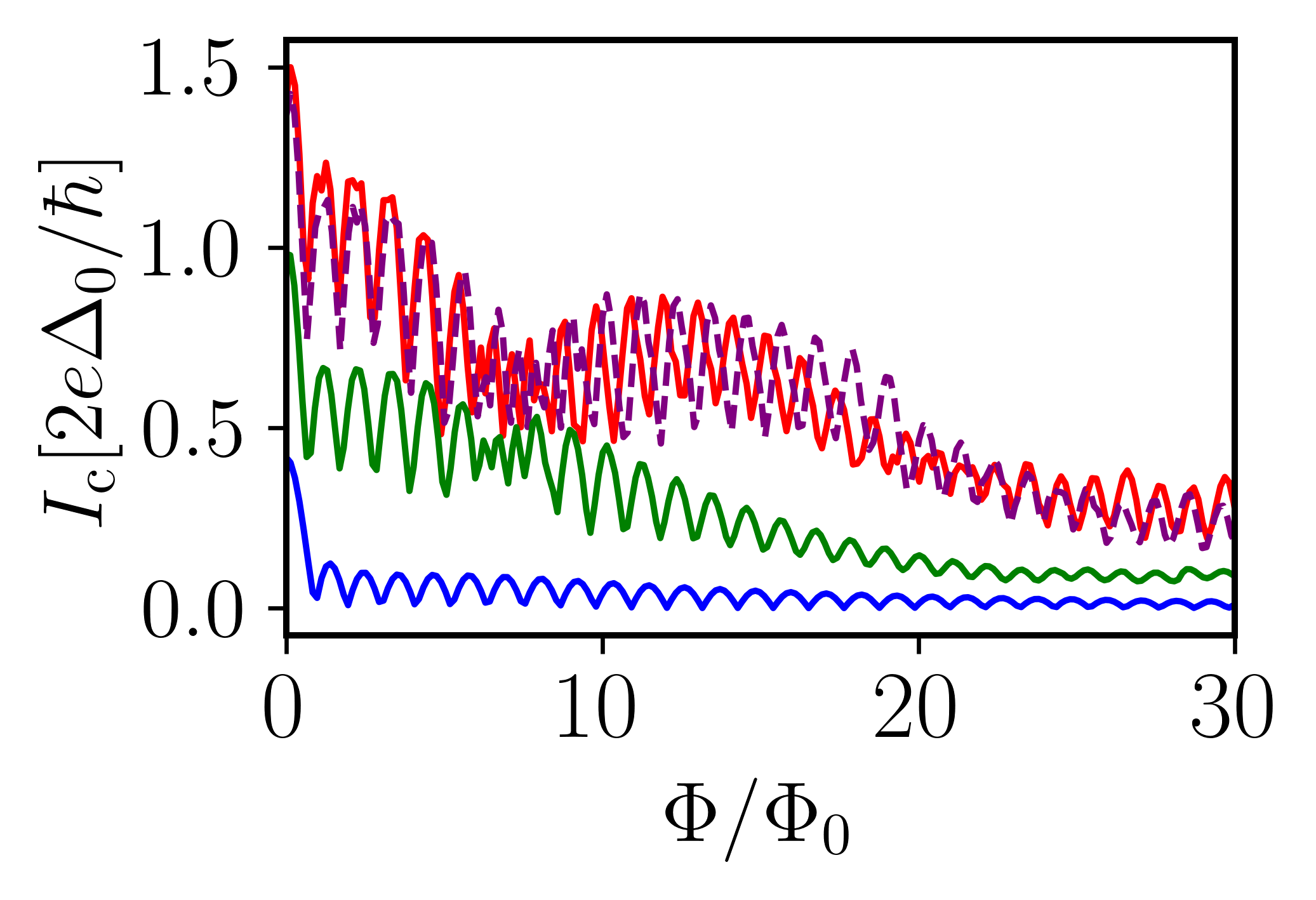}\put(2,70){a)}\end{overpic}\hspace{2em}
\begin{overpic}[width=0.35\linewidth]{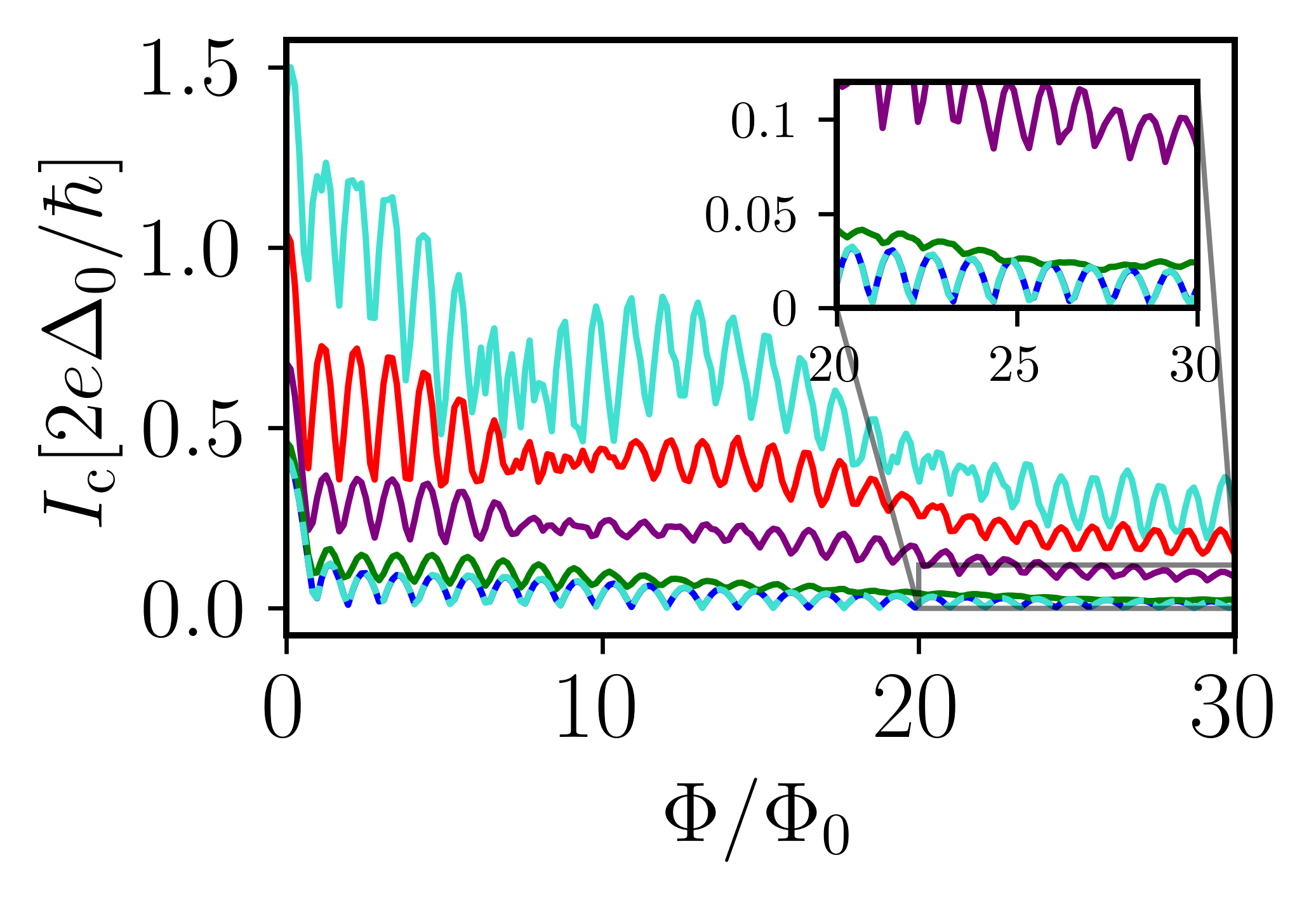}\put(2,70){b)}\end{overpic}

\caption{Critical supercurrent $I_{\rm c}$ (in units of $2e\Delta_0/\hbar$) as a function of the magnetic flux $\Phi$ (in units of the superconducting magnetic flux quantum $\Phi_0 = h/2e$), produced by a perpendicular magnetic field threading the junction. The parameters used are $\lambda_{\rm R} = 5~ \hbar v/L$, $\lambda_{\rm VZ} = \hbar v /L$, $\mu = 0$ and a junction length and width of $L=59a$ and $W=5L$. In a) we show the clean edge case with different edge terminations; in red the results for a pure zigzag ribbon with $(m,n)=(1,0)$, in purple (green) those for a $(m,n)=(2,1)$ ($(m,n)=(1,2)$) termination, and in blue the ones for a pure armchair ribbon, $(m,n)=(0,1)$. In b) we show the same plot as in a) but for junction with disordered edges; in particular, in red (blue) we show the results for a zigzag (armchair) edge with small disorder, in purple (green) those for a zigzag (armchair) edge with heavy disorder, in solid (dashed) cyan we plot the clean zigzag (armchair) case for reference. The inset in b) shows a magnified view of the high magnetic field region highlighting the results for heavily disordered edges.}
\label{fig:MagneticInterference_HighFields}
\end{figure*}

In this section, we analyze the magnetic interference properties of the \gls*{gjj} when the scattering graphene layer is in the \gls*{qvsh} phase and is threaded by an external static magnetic field applied perpendicular to it.
Specifically, we take into account only the orbital effects of the magnetic field, and we study the robustness of the critical current at high magnetic fluxes and the non-reciprocity of transport at low magnetic fluxes. We refer to Appendix \ref{sec:App:simulations_details} for more details on the implementation of the magnetic field in the numerical calculations.

In general, superconducting transport in a \gls*{jj} can be composed of delocalized contributions, which are spread along the width of the junction, and more localized ones, which are instead highly spatially confined.
In the former case, the supercurrent contribution is sensitive to the spatial variations of the gauge-invariant phase difference, resulting in a contribution that averages out for high magnetic fluxes \cite{Tinkham_1996_a}.
An example of this is the typical Fraunhofer pattern in the critical current of a tunnel \glspl*{jj}, $I_{\rm c}(\Phi) = I_{\rm c}(0) | \sin (\pi \Phi/\Phi_0) / (\pi \Phi/\Phi_0)|$, arising from low-transmission and spatially homogeneous bulk contributions.
An example of the latter is that of two $\delta$-localized contributions of unit transmission at the edges of the junction; because of the extreme localization, these are insensitive to the gauge-invariant phase variation and produce a periodic critical current as \cite{Beenakker_2015_a}
\begin{equation}\label{eq:Beenakker_Interference}
    I_{\rm c}(\Phi) = (I_{\rm c}(0)/2)(1+|\cos(\pi \Phi/\Phi_0)|)~,    
\end{equation}
which persists indefinitely in the magnetic field strength.
Therefore, the persistence of the critical current at high magnetic fields is an indicator of the degree of localization of the supercurrent.


\subsection{Supercurrent robustness at high magnetic fluxes}\label{sec:SupercurrentRobustness}

As discussed in previous sections, in a proximitized \gls*{gjj} with graphene in the \gls*{qvsh} phase, the transport phenomenology is rich.
In particular, when the Fermi level lies within the band gap, we generally have both extended contributions from evanescent bulk states and localized contributions from edge states, with different localizations and transmission probabilities.

The resulting interference pattern for the critical supercurrent, defined in Eq.~\eqref{eq:CriticalCurrent_Definition} and computed at zero Fermi level (charge neutrality point), is shown in Fig.~\ref{fig:MagneticInterference_HighFields}~a), where the different edge terminations have the same color code as in Fig.~\ref{fig:Crit_Current_R5S1_CleanAndDisordered}.
The \gls*{soc} parameters and junction dimensions are the same as in the previous sections.
From Fig.~\ref{fig:MagneticInterference_HighFields}~a), we can make a few observations.
First, we see that, in the case of an armchair termination, the lack of edge states, and therefore of localized currents, leads to a complete suppression of the residual critical current at high magnetic fields, in agreement with the \gls*{qvsh} model and with our earlier results on superconducting transport in zero external magnetic field.
Interestingly, in this case, the interference pattern slightly differs from the standard Fraunhofer form in both the frequency of the nodes, which is lower, and the overall decay is more gradual. This behavior arises from the spatial profile of the supercurrent in the junction, which is spatially spread out but not exactly uniform; further details about it are provided in Appendix~\ref{sec:App:CleanArmchair_CurrentPlot}.

Another feature that we observe is that all non-armchair edge terminations show a beat around $\Phi/\Phi_0 \approx 7$.
This is the result of two oscillating patterns of the kind we saw in Eq.~\eqref{eq:Beenakker_Interference} with slightly different frequencies: one frequency is due to the pseudohelical edge states, which are strongly localized at the sample boundaries and therefore enclose nearly the entire area of the junction. 
The second frequency originates from the valley edge states, which, being more spatially extended \cite{Frank_2018_a}, enclose a smaller effective area and consequently acquire a different magnetic phase.
With increasing magnetic flux, the contribution to the total critical current of the valley edge states averages off, and for $\Phi/\Phi_0 \gtrsim 20$, most of the remaining critical current is due only to the pseudohelical edge states.
Lastly, we observe that with increasing armchair content in the edge termination, the residual critical current diminishes, again consistent with the less localized nature of edge states \cite{Akhmerov_2008_a}.

Moreover, we study the magnetic interference pattern for disordered edge terminations.
We again distinguish small and heavy disorder, which are implemented in the same way as before.
The results are shown in Fig.~\ref{fig:MagneticInterference_HighFields}~b).
In solid (dashed) cyan we plot the same results as in a) for a clean zigzag (armchair) termination as a reference.
In red (blue) we show the results for small disorder in a zigzag (armchair) edge, and in purple (green) the same ones for heavy disorder.
We observe that, for small disorder, the interference pattern remains almost unaffected in the zigzag configuration. In fact, it retains the same main characteristics as in the clean case, with only a modest reduction of the residual critical current, corresponding to the overall reduction in transmission.
For the armchair termination, a small disorder has essentially no effect.
For heavy disorder, we find that the zigzag termination follows the same trend: it exhibits the same features, but with a further lowering of the residual critical current. Again, we attribute this to a further reduction in transmission, together with an increased localization length of the edge states contributing to the transport.
For the armchair termination, instead, the situation is the opposite.
We find that heavy disorder actually increases both the supercurrent at low magnetic fluxes and the residual critical current at high magnetic fluxes, as can be seen in the inset of Fig.~\ref{fig:MagneticInterference_HighFields}~b).
This means that because of edge disorder, more of the transport occurs in a localized fashion near the edges of the sample.
All of these findings are consistent with the results of the previous section and reflect the localized nature of the supercurrent in \gls*{qvsh} \gls*{gjj}.

\subsection{Superconducting diode effect}




%
\begin{figure*}[t!]
\centering

\begin{overpic}[width=0.6\linewidth]{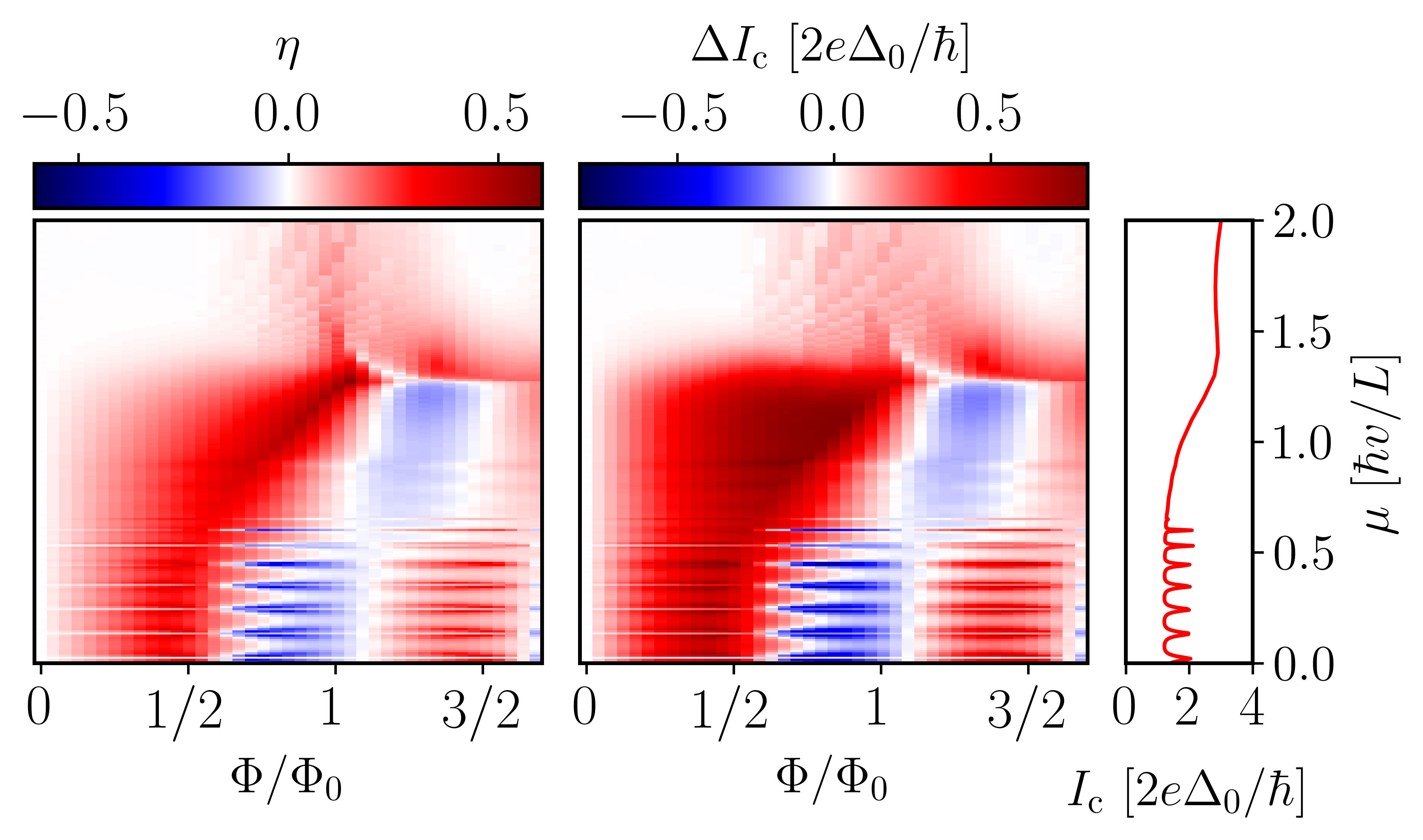}\put(0,55){a)}\put(38,55){b)}\put(78,46){c)}\end{overpic}\hspace{2em}
\begin{overpic}[width=0.25\linewidth]{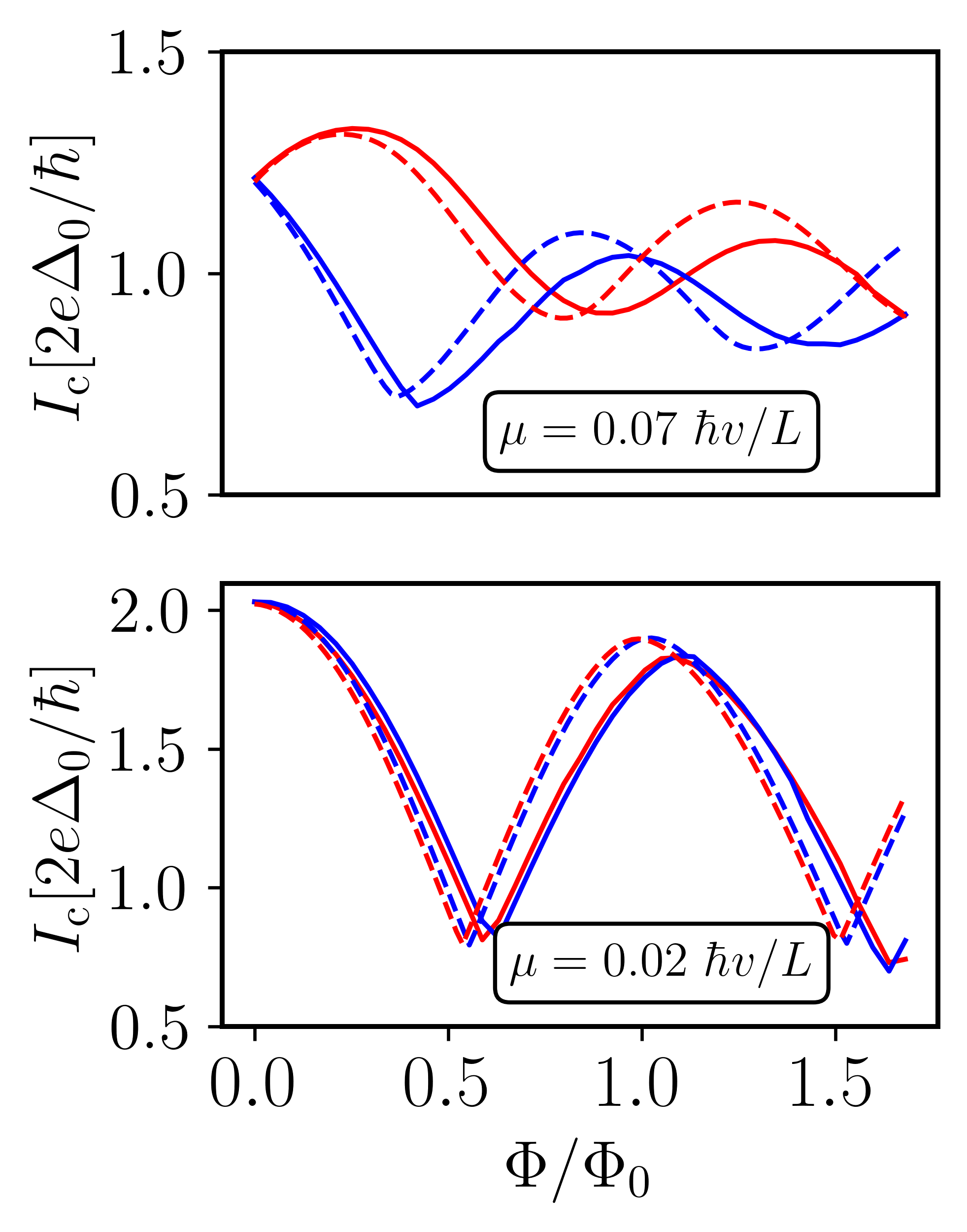}\put(-2,100){d)}\put(-2,55){e)}\end{overpic}

\caption{a) Efficiency $\eta$ and b) non-reciprocal supercurrent $\Delta I_c$ (in units of $2 e \Delta_0/\hbar$) for a \gls*{gjj} in the \gls*{qvsh} phase with zigzag edge terminations, as functions of the Fermi level of the graphene scattering region $\mu$ (in units of $\hbar v/L$) (vertical axis) and the magnetic flux $\Phi$ (in units of the superconducting
magnetic flux quantum $\Phi_0=\hbar/(2e)$) threading the junction (horizontal axis).
As a reference, in c) we show the critical current $I_{\rm c}$ (horizontal axis) as a function of the Fermi level $\mu$ (vertical axis) for a junction with zigzag edges at zero magnetic flux, the same result shown in Fig.~\ref{fig:Crit_Current_R5S1_CleanAndDisordered}~a).
Non-reciprocal critical currents d) for small $\tau_{\rm T} \approx 0.25$ and e) for high $\tau_{\rm T} \approx 1$, at the indicated Fermi level, $\mu$: in red (blue) we show $\max_{\varphi} I(\varphi)$ ($-\min_{\varphi} I(\varphi)$);
solid lines are calculated numerically, while dashed lines are computed from the minimal analytical description of Eq.~\eqref{eq:MinimalSDEModel}, as explained in the main text.
In all panels, the \gls*{soc} parameters used are the same as in the previous sections: $\lambda_{\rm R} = 5~ \hbar v/L$, $\lambda_{\rm VZ} = \hbar v /L$ with a junction length and width of $L=59a$ and $W=5L$.}
\label{fig:SDE_EfficiencyAbsDiffCritCurr_WithCuts}
\end{figure*}
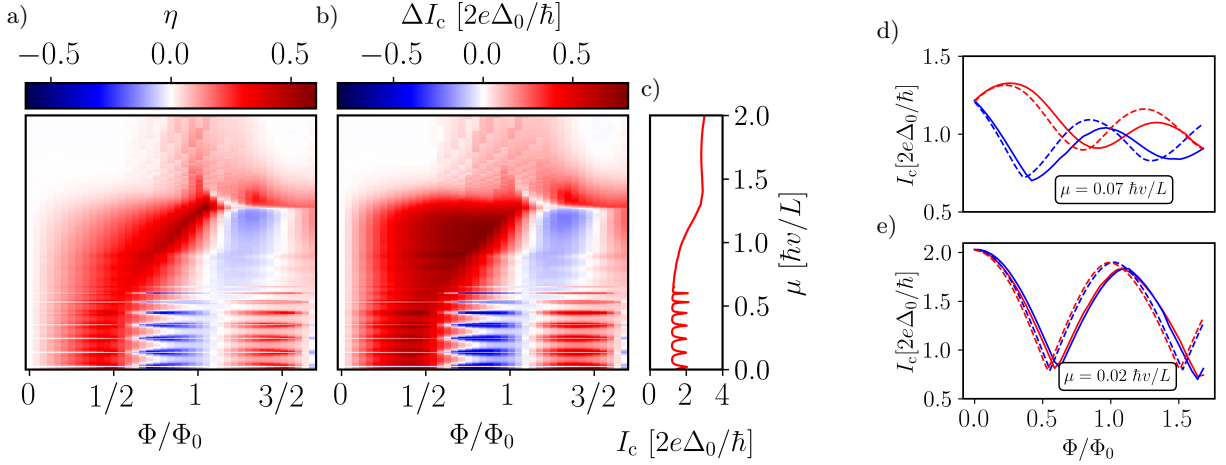
%

Here, we continue our investigation of the junction under an applied magnetic field, focusing on the emergence of non-reciprocal transport.
\glsreset{sde}
The non-reciprocity of the supercurrent is known as \gls*{sde} \cite{Ando_2020_a} and the two main quantities that are used to characterize it are the \textit{non-reciprocal supercurrent}, $\Delta I_{\rm c}$, defined as the difference between the maximum supercurrents flowing in opposite directions, and the \textit{diode efficiency}, $\eta$, which characterizes the relative asymmetry of the superconducting diode.
In a \gls*{jj}, they can be expressed in terms of the critical currents as \cite{Baumgartner_2022_b,Jeon_2022_a,Turini_2022_a,Fu_2024_a}
\begin{align}\label{GJJE:eq:NonReciprocalSupercurrent}
    \Delta I_{\rm c} & = \max_{\varphi} I(\varphi) + \min_{\varphi} I(\varphi), \\
    \eta & = \frac{\Delta I_{\rm c}}{\max_{\varphi} I(\varphi) -\min_{\varphi} I(\varphi)}.
\end{align}

In Fig.~\ref{fig:SDE_EfficiencyAbsDiffCritCurr_WithCuts}~a-b), we show the results of $\eta$ and $\Delta I_{\rm c}$, as functions of the magnetic flux, $\Phi$, and the Fermi level, $\mu$, for a \gls*{gjj} with graphene in the \gls*{qvsh} phase and zigzag edge terminations.
We find a high degree of non-reciprocity with maximum efficiencies reaching close to $\eta \approx 0.6$, while still maintaining a non-reciprocal supercurrent comparable to the values of the critical current at zero magnetic flux.
Moreover, the \gls*{sde} behavior of the junction is limited to Fermi level values $\mu$ close to the energy gap and to zigzag-like terminations (the armchair one shows vanishing non-reciprocal transport as shown in Appendix \ref{sec:App:SDE_NonZigzag}).
We also find that there are white horizontal stripes with zero efficiency that perfectly align with the constructive interference peaks of the critical current analyzed earlier and shown again in Fig.~\ref{fig:SDE_EfficiencyAbsDiffCritCurr_WithCuts}~c).
%

We have previously observed that when the current is injected from pristine to zigzag \gls*{qvsh} graphene, the resulting transport becomes asymmetric, with one edge being more transmissive than the other.
For small magnetic fluxes threading the junction, $\Phi/\Phi_0 \ll W/\ell_i$, where $\ell_i$ are the localization lengths of the edge states, we are justified in neglecting the spatial extent of the edge states and adopting a minimal description given by
\begin{equation}\label{eq:MinimalSDEModel}
    I(\varphi,\Phi) = I_{\rm B} (\varphi,\Phi) + I_{\rm E} (\varphi,\Phi),
\end{equation}
with \cite{Tinkham_1996_a}
%
%
%
%
\begin{widetext}
\begin{gather}
    I_{\rm E} (\varphi,\Phi) = \frac{2e \Delta_0}{\hbar} \left[ \frac{\tau_{\rm B} \sin (\varphi - \pi \Phi/\Phi_0)}{2 \sqrt{1-\tau_{\rm B} \sin^2 ((\varphi - \pi \Phi/\Phi_0)/2)}} + \frac{\tau_{\rm T} \sin (\varphi + \pi \Phi/\Phi_0)}{2 \sqrt{1-\tau_{\rm T} \sin^2 ((\varphi + \pi \Phi/\Phi_0)/2)}} \right], \label{eq:MinMod_Edge} \\
    I_{\rm B} (\varphi,\Phi) = \frac{e\Delta_0}{2\hbar} \left(\sum_{\tau \neq \{\tau_{\rm T},\tau_{\rm B}\}} \tau \right)\frac{\sin(\pi \Phi/\Phi_0)}{\pi \Phi/\Phi_0} \sin(\varphi)~, \label{eq:MinMod_Bulk}
\end{gather}
\end{widetext}
where the first equation accounts for the localized edge contributions coming from the bottom and top edges, respectively, and the second equation accounts for the delocalized bulk contributions.
All the transmission probabilities used in the minimal model are obtained by solving the scattering problem for a non-superconducting junction at zero magnetic flux, as described in Eq.~\eqref{eq:Supercurrent_BeenakkerFinal}.
For $\mu \geq 0$, we find $\tau_{\rm B} \approx 1$, while $\tau_{\rm T}$ ranges from as low as $0.25$ to $1$ depending on the degree of constructive interference.
When $\tau_{\rm T}$ is small, its channel contribution can be approximated with a sinusoidal expression, and the edge contribution of Eq.~\eqref{eq:MinMod_Edge} now mixes two terms in the current-phase relation of the junction as
\begin{equation}\label{eq:MinimalSDELowT}
\begin{aligned}
    I_{\rm E} \left( \varphi, \Phi \right) & = \frac{e \Delta_0}{\hbar} \left\{ \sin\left(\frac{\varphi}{2} + \frac{\pi \Phi}{2\Phi_0}\right) \text{sgn} \left[ \cos \left( \frac{\varphi}{2} + \frac{\pi \Phi}{2\Phi_0} \right) \right] \right. \\
    & \left. + \frac{\tau_{\rm T}}{2} \sin \left(\varphi - \pi \frac{\Phi}{\Phi_0}\right) \right\},
\end{aligned}
\end{equation}
which was recently proposed in the literature as a general way to generate non-reciprocal supercurrent in Josephson interferometers \cite{Souto_2022_a}.
In our system, this is realized in a single device.
When, instead, $\tau_{\rm T}$ approaches perfect transmission, the total edge contribution returns to the periodic behavior of Eq.~\eqref{eq:Beenakker_Interference}, and the asymmetric component of the transport vanishes.
We compare the minimal description given by Eq.~\eqref{eq:MinimalSDEModel} with the results obtained numerically in Fig.~\ref{fig:SDE_EfficiencyAbsDiffCritCurr_WithCuts}~d), for small $\tau_{\rm T} \approx 0.25 $, and in Fig.~\ref{fig:SDE_EfficiencyAbsDiffCritCurr_WithCuts}~e), for high $\tau_{\rm T} \approx 1 $.
Specifically, solid lines are the numerical results and dashed lines refer to the minimal analytical description: in red (blue) we show $\max_{\varphi} I(\varphi)$ (-$\min_{\varphi} I(\varphi)$).
We observe that the minimal model closely matches the numerical results; the only discrepancies concern the oscillation frequency and a slight damping of the critical current, both of which can be attributed to the neglect of the edge states’ finite localization length.
In this way, we identify the asymmetric edge transport in the junction as the source of the observed \gls*{sde}.

We note that the \gls*{sde} observed in our setup is different from the one observed in the literature of planar \gls*{jj} with \gls*{rb} coupling \cite{Turini_2022_a,Nadeem_2023_a,Fu_2024_a,Bhowmik_2025_a}.
In fact, in those systems, the \gls*{sde} is observed when the system is coupled to a Zeeman splitting due to a magnetic field parallel to the junction.
In our system, instead, the \gls*{sde} originates purely from the orbital effect of the magnetic field, since we have neglected any Zeeman coupling in the junction.
It is a behavior similar to the general theory developed in Ref.~\cite{Chirolli_2025_a}, in which the \gls*{sde} originates from a mirror asymmetry along the width of the scattering region. 


\section{Conclusions}\label{sec:conclusions}

In this work, we have investigated the normal and superconducting transport properties of spin-orbit coupled graphene via proximity effect, focusing on the edge contribution.
We showed that both helical and pseudohelical edge states act as efficient valley filters for bulk electrons.
In the \gls*{qvsh} phase, the resulting valley polarization combines with the mirror-asymmetric structure of the zigzag nanoribbons to produce a strong asymmetry in transport along the two edges of the junction.
Having established the origin of the asymmetric normal transport, we then studied its consequences on superconducting transport in a \gls*{qvsh} \gls*{gjj}.
By analyzing the critical current as a function of chemical potential, we found that, as expected, transport inside the bulk gap is dominated by edge states.
In this regime, the critical current exhibits resonances, which are associated with the weakly transmitting edge channel and the coherent nature of the setup.
We find that the critical current is robust for different ribbon orientations, as long as the termination contains a zigzag component, and against small disorder in the form of edge roughness.
Stronger roughness markedly modifies edge conduction and can generate edge-state transport even for armchair terminations, where it does not occur in the clean limit.
Despite these quantitative changes, the edge contribution to the supercurrent remains clearly visible, demonstrating the robustness of this transport mechanism.

The edge-dominated nature of the superconducting transport is further confirmed by the magnetic response of the junction. The magnetic interference pattern remains visible over a broad range of magnetic fields and displays slowly damped periodic oscillations, a characteristic signature of localized edge transport. The same behavior persists in the presence of moderate edge disorder, reinforcing the picture obtained from the critical-current analysis.

Finally, we examine the regime of low magnetic field and show that the asymmetric edge transport gives rise to a substantial \gls*{jde}.
For zigzag junctions, the diode efficiency can reach values of approximately $60$\%.
Interestingly, within the bulk gap, the diode efficiency exhibits an oscillatory dependence on the chemical potential that follows the transmission resonances of the edge states, providing a direct link between the nonreciprocal supercurrent and the underlying edge transport.
Together with our minimal edge-state model, these results identify the asymmetry between the two edge channels as the origin of the diode effect.

Overall, our results show that a \gls*{gjj} made with \gls*{qvsh} graphene provides an interesting platform in which valley filtering, edge-state transport, and spin-orbit coupling combine to generate robust superconducting transport and strong supercurrent non-reciprocity under weak orbital magnetic fields, with potential applications in superconducting quantum circuitry.

{\it Note added.} As we were finalizing this manuscript, we became aware of the work by Villani \textit{et al.}, who reported a Josephson diode effect in graphene/hBN Josephson junctions. Their mechanism does not rely on spin-orbit coupling and is instead attributed to the combined effect of an out-of-plane magnetic field and mirror-symmetry breaking induced by asymmetric long-range disorder~\cite{villani_preprint_2026}.

%

\begin{acknowledgments}
	The authors thank G.G.N. Angilella, L. Giannelli, V. Varrica, for their insightful comments and constructive feedback throughout various stages of this work.
    F.B. and E.P. thank the PNRR MUR project PE0000023-NQSTI.
    E.P. acknowledges support from COST Action CA21144 superqumap. 
    F.M.D.P. acknowledges support from the project PRIN 2022 - 2022XK5CPX (PE3) SoS-QuBa - ``Solid State Quantum Batteries: Characterization and Optimization". 
    G.F. thanks for the support ICSC - Centro Nazionale di Ricerca in High-Performance Computing, Big Data and Quantum Computing under project E63C22001000006, and Universit\`a degli Studi di Catania, 
    project TCMQI PIACERI 2024/2026.
    F.B. and F.M.D.P. acknowledge support from  Centro Siciliano di Fisica Nucleare e Struttura della Materia (CSFNSM).
\end{acknowledgments}

\bibliographystyle{mprsty}
\bibliography{Bibliography}

\appendix
\numberwithin{equation}{section}
\renewcommand\thefigure{\thesection.\arabic{figure}}
\setcounter{figure}{0}

\section{Details of the numerical calculations}\label{sec:App:simulations_details}

We aim to simulate a \gls*{gjj} with a length of around $200$ nm and a width to length ratio of $W/L = 5$.
In recent transport measurements for graphene on transition metal dichalcogenides, in particular on WS$_2$, junctions of similar lengths were found to be in the short junction limit, while still behaving ballistically \cite{Wakamura_2020_a}.
When considering graphene in the \gls*{qvsh} phase, from Sec.~\ref{sec:QVSH_AsymmetricTransport} onward, the \gls*{soc} energies of $\lambda_{\rm VZ}$ and $\lambda_{\rm R}$ that we consider are in the upper limits of what was found experimentally in Refs.~\cite{Wang_2016_a,Wang_2019_c,Island_2019_a,Sun_2023_a}, namely $\lambda_{\rm VZ} \approx 3~\text{meV}$ and $\lambda_{\rm R} \approx 15~\text{meV}$; while for graphene in the \gls*{qsh} phase, studied in Sec.~\ref{sec:App:QSH_DoubleJunction}, we take, for simplicity, $\lambda_{\rm K} = \lambda_{\rm VZ}$, which produces a similar edge states energy dispersion to the pseudohelical ones.

To reduce the computational resources needed, we use a scaling procedure that simulates the junction using a larger lattice constant $a s_{\rm f}$, where $s_{\rm f}$ is the scaling factor, with a renormalized hopping parameter $t/s_{\rm f}$ to make sure that the intrinsic graphene energy spectrum, given by $E_0 = (\sqrt{3}/2)at|\bm{k}|$, is unchanged \cite{Liu_2015_a}.
The validity criterion for the scaling procedure is given by $s_{\rm f} \ll 3t\pi/|E_{\rm max}|$, where $|E_{\rm max}|$ is the maximum energy of interest that needs to be investigated \cite{Liu_2015_a}.
So, taking $\lambda_{\rm R} \approx 15~\text{meV}$, the validity criterion in our case becomes $s_{\rm f} \ll \frac{3t\pi}{\lambda_{\rm R}} \approx \frac{3\cdot 2.8\cdot\pi}{15\cdot 10^{-3}} \approx 1760$. For our calculations, we take $s_{\rm f} \approx 13.7$, which corresponds to a scaled junction length of $L= 59a$ and a width of $W = 5L = 295a$.
Operatively, we do the equivalent way of keeping $t$ numerically constant and scaling the \gls*{soc} energies by the $s_f$ factor; for better clarity, all the energies are then indicated in the Thouless energy scale $\hbar v /L$.

\begin{figure}
\centering
\begin{overpic}[width=0.75\columnwidth]{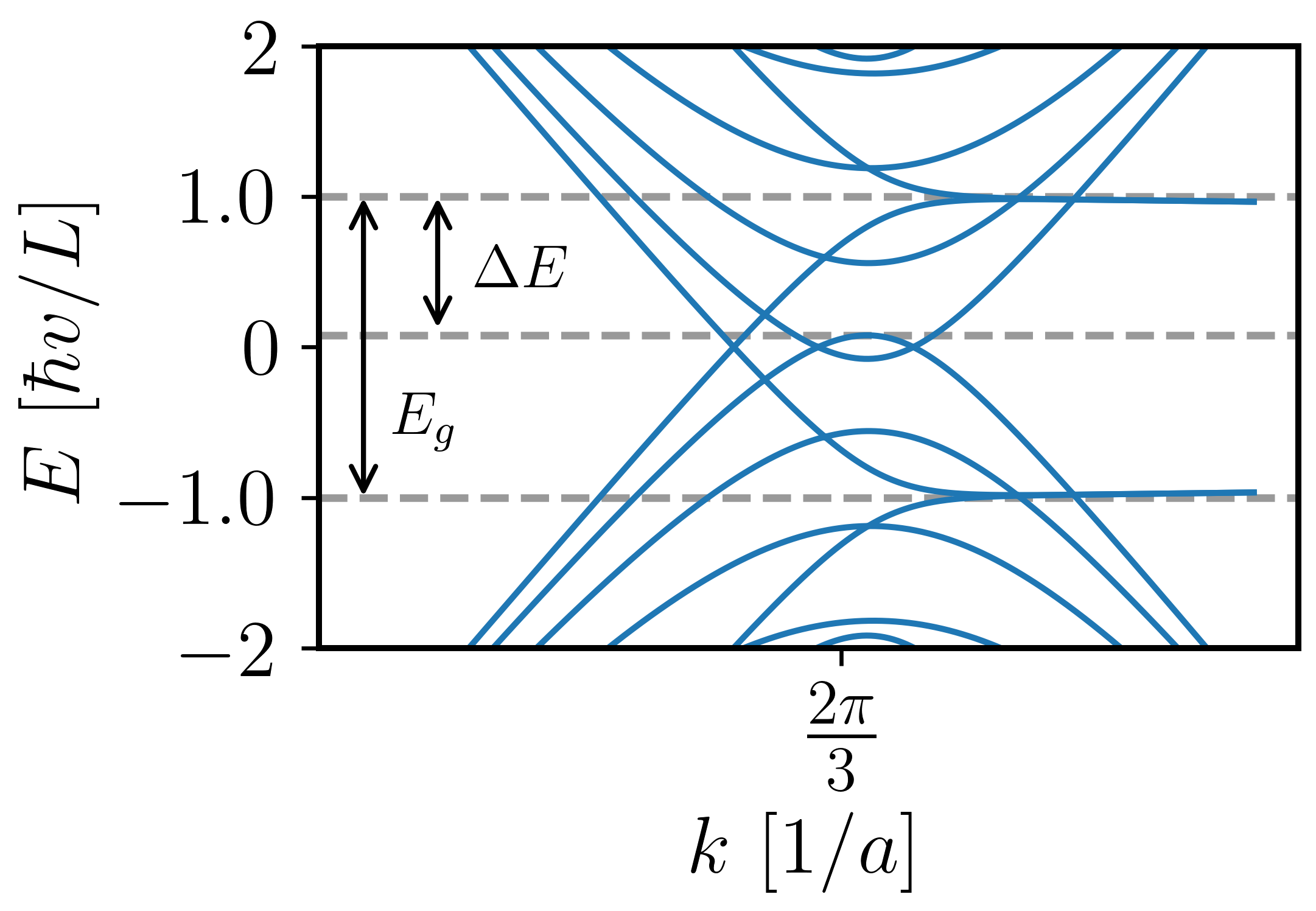}\end{overpic}
\caption{Zoom around the $K$ valley of the energy band structure for a \gls*{qvsh} zigzag ribbon of width $W=295a$ and $\lambda_{\rm VZ} = \hbar v /L$, with $L$ the junction's length of $L = W/5$ considered in the main text.
$E_g = \lambda_{\rm VZ}$ indicates the inverted gap and $\Delta E$ the finite-size level spacing.}
\label{fig:App:ZeroRashba_EnergyBands}
\end{figure}

In Fig.~\ref{fig:App:ZeroRashba_EnergyBands} we show a zoom around the $K$ valley of the band structure of a zigzag ribbon of width $W=295a$ and $\lambda_{\rm VZ} = \hbar v /L$, $\lambda_{\rm R} = 0$.
We see that
\begin{equation}\label{eq:App:Condition_QVSH}
    \Delta E < E_g/2,
\end{equation}
where $E_g = 2\lambda_{\rm VZ}$ represents the inverted gap, and $\Delta E$ the finite-size level spacing.
Hence, according to Ref.~\cite{Frank_2018_a}, we are just within the condition for the existence of the valley-edge states in the \gls*{qvsh} graphene phase.
Although, we consider here a different limit from Ref.~\cite{Frank_2018_a} with $\lambda_{\rm R} > \lambda_{\rm VZ}$, which, according to our numerical calculations, helps to stabilize the existence of the valley-edge states to produce the band structure shown in Fig.~\ref{fig:JP_EC_VZ}.

The junction's sectors composed of pristine graphene (without spin orbit couplings) are considered in the high doping limit. Numerically, we set the Fermi level $\mu_R = 0.2t$, in such a way as to be much higher than the energy scales used in the scattering region of the junction, but still in the linear regime of the graphene Dirac cones, so we can neglect trigonal warping effects \cite{CastroNeto_2009_a}.

In Sec.~\ref{sec:MIP}, we study the interference pattern of the supercurrent when the \gls*{gjj} is threaded by a perpendicular magnetic field.
With a junction of total surface area of $A_0 \approx 0.2~\mu$m$^2$ the interference pattern arises with magnetic fields of the order of $B_{\Phi_0} = \Phi_0/A_0 \approx 10 \text{ mT}$, where $\Phi_0 = h/2e$ is the superconducting magnetic flux quantum.
The Zeeman splitting energy for these magnetic fields is $E_{\rm Z} \approx 2\mu_{\rm B} B_{\Phi_0} \approx 0.6$ $\mu$eV, which is far below any other energy scale we have considered.
For this reason, we have included only the orbital effects of the magnetic field in the tight-binding calculations while neglecting the Zeeman coupling.
The orbital effects are included in the tight-binding calculations via the Peierls substitution~\cite{Grosso_2000_a}, which changes the hopping from site $i$ to site $j$ as
\begin{equation}\label{eq:App:Periels_Substitution}
    t_{ij} \longrightarrow t_{ij} \exp\left\{-i \frac{e}{\hbar} \int_i^j \bm{A}\cdot d\bm{l}\right\}.
\end{equation}
For a junction along the $x$ horizontal direction, it is useful to employ the Landau gauge
\begin{equation}\label{eq:App:Landau_Gauge}
    \bm{A} = (B_z y,0,0)
\end{equation}
to describe a perpendicular magnetic field $\bm{B} = (0,0,B_z)$, which substituted into Eq.~\eqref{eq:App:Periels_Substitution} gives us
\begin{equation}\label{eq:App:Periels_Substitution_Gauge}
    t_{ij} \longrightarrow t_{ij} \exp\left\{i \pi \frac{\Phi}{\Phi_0} \frac{y_j + y_i}{2} (x_j - x_i)\right\},
\end{equation}
where $\Phi = B_z (LW) = B_z A_0$ is the magnetic flux threading the junction and $x$ ($y$) is measured in units of $L$ ($W$).
Moreover, the scaling parameter now also has to satisfy $s_{\rm f} \ll l_B/a \approx 5700/\sqrt{B_z [\text{mT}]}$, where $l_B = \sqrt{\hbar/eB_z}$ is the magnetic length, in order for the Peierls substitution to be valid \cite{Goerbig_2011_a,Liu_2015_a}.
Even for high magnetic fluxes of $\Phi = 30 \Phi_0$, we are a full order of magnitude within the validity of the scaling, $s_{\rm f} \approx 13.7 \ll 323$, which gets much better for smaller magnetic fluxes.

\begin{figure}
\centering
\begin{overpic}[width=0.5\columnwidth,trim={0 0cm 0 0cm}]{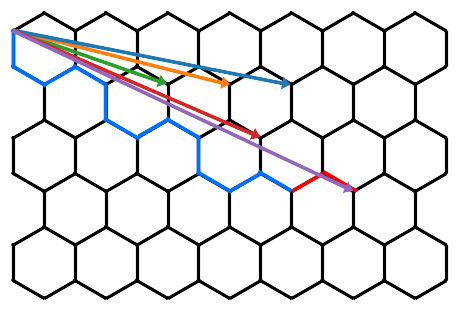}\end{overpic}
\caption{Graphene honeycomb lattice on top of which we plot the vectors that define the periodicity of a given edge termination.
Starting from blue and proceeding clockwise to purple, the edge termination has increasing armchair content, in order: $(m,n)=(3,1),(2,1),(1,1),(1,2),(1,3)$, where the first (second) value indicates the number of zigzag (armchair) sections in the periodicity vector.
For the $(m,n)=(1,3)$ edge termination (purple arrow), a representative boundary cell is highlighted. Its armchair and zigzag segments are shown in light blue and bright red, respectively.
}
\label{fig:App:Edge_Terminations}
\end{figure}

Finally, in this work, we define each edge termination by two numbers $(m,n)$ where $m$ ($n$) indicates the number of zigzag (armchair) sections in the periodicity vector.
Figure \ref{fig:App:Edge_Terminations} shows the periodicity vectors that define some edge terminations we considered: starting from blue and proceeding clockwise to purple, the edge termination has increasing armchair content, in order: $(m,n) = (3,1),(2,1),(1,1),(1,2),(1,3)$.
Note that different edge terminations have periodicity vectors of different lengths.

\section{QSH double junction as valley-filter}\label{sec:App:QSH_DoubleJunction}

\begin{figure*}[t!]
\centering
\begin{overpic}[width=0.30\linewidth]{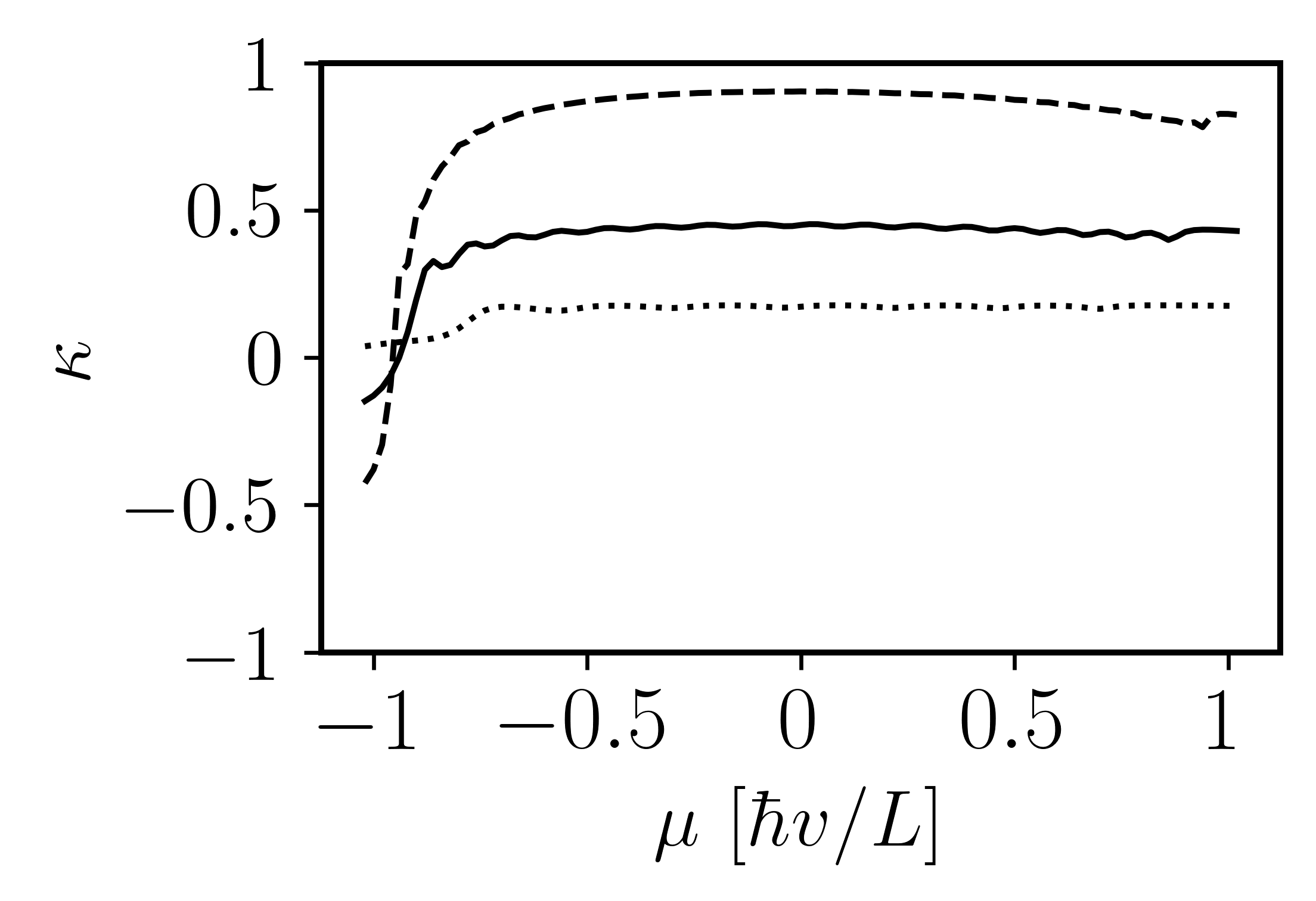}\put(0,70){a)}\end{overpic}
\begin{overpic}[width=0.32\linewidth]{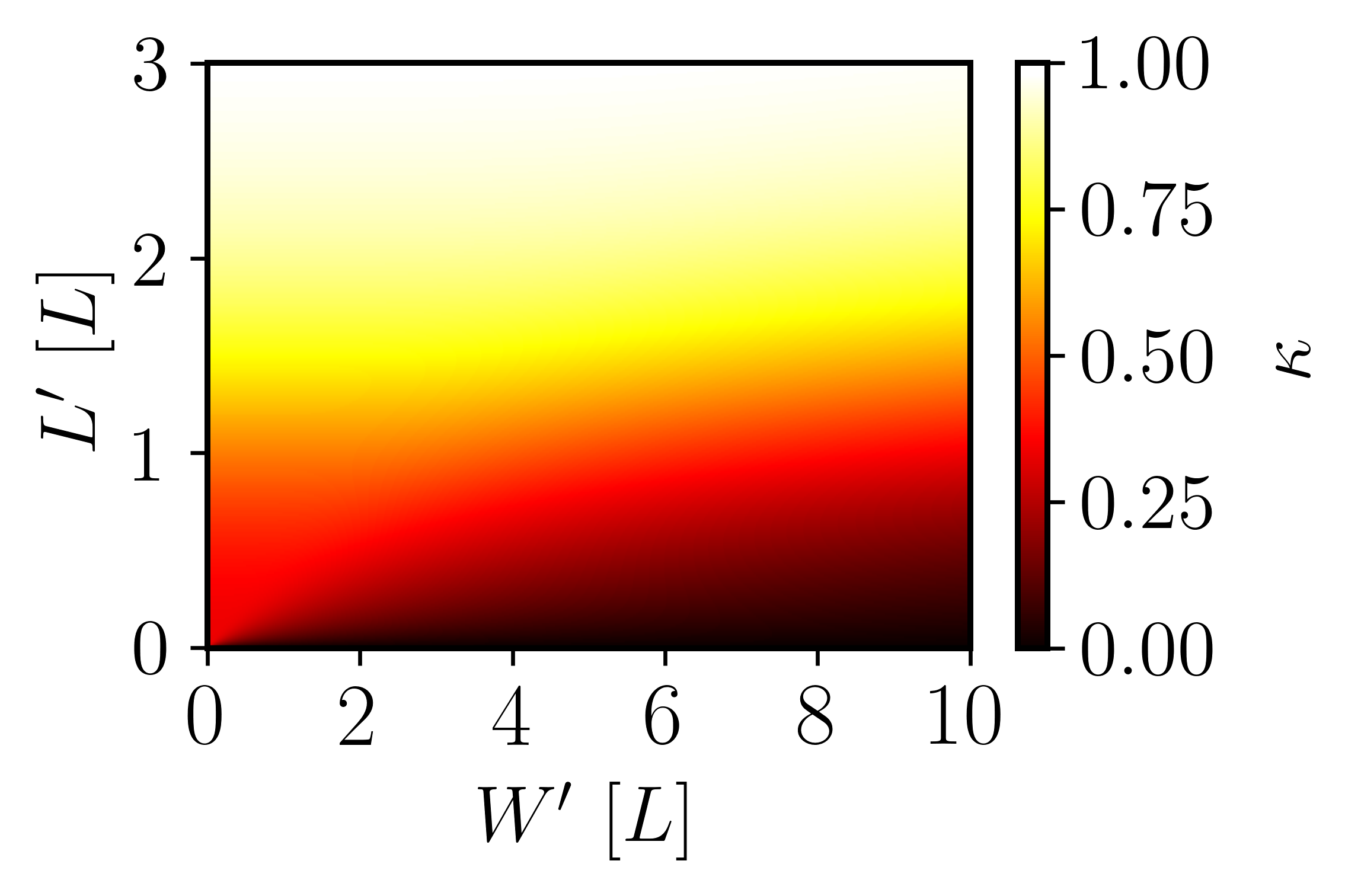}\put(0,65.5){b)}\end{overpic}
\begin{overpic}[width=0.3\linewidth]{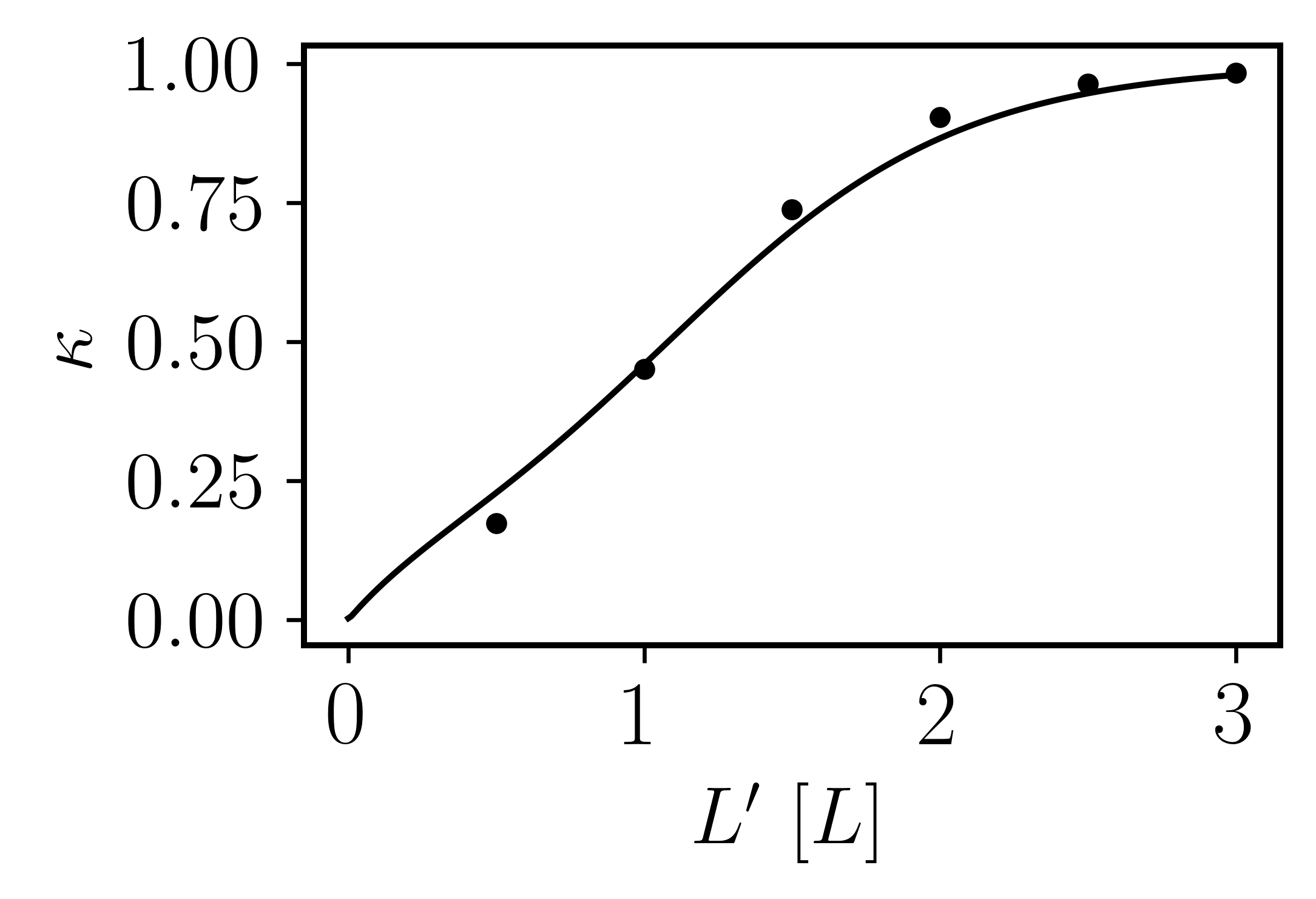}\put(0,70){c)}\end{overpic}

\caption{a) Valley polarization of the incoming current, $\kappa$, as defined in the main text, as a function of Fermi level, $\mu$, calculated numerically for different lengths of the scattering region: $L' = L/2, L, 2L$ in dotted, solid, and dashed, respectively.
b) Analytically computed valley polarization, $\kappa$, at zero Fermi level, $\mu = 0$, for varying junction length and width, in the limit $\mu_{\rm R} \to \infty$.
c) Vertical cut along $W'=5L$ of b): the analytical results match well with the numerical ones, shown as black dots, taken from a) at $\mu = 0$.
In all panels the strength of the \gls*{soc} interaction is fixed at $\lambda_{\rm KM} = \hbar v /L$.}

\label{fig:App:VP_KM_DoubleJunction}
\end{figure*}

Here, we show the results for the valley-filtering effect in a double junction made with \gls*{qsh} graphene.
We consider the current injected from a right pristine (without spin-orbit couplings) graphene lead that scatters through a \gls*{qsh} region and is then transmitted again to a left pristine graphene lead.
Specifically, we consider the same junction parameters as in the main text: $\lambda_{\rm KM} = \hbar v /L$ with $L=59a$ and $W=5L$ for the \gls*{qsh} scattering region, and $\mu_{\rm R} = 0.2t$ for the pristine leads.
We compute the incoming valley polarization, which is equal to the outgoing one, in the same way as done in the main text, through Eqs.~(\ref{eq:TransmisssionEigenproblem_Original}-\ref{eq:IncomingValleyPolarization}).
The results are shown in Fig.~\ref{fig:App:VP_KM_DoubleJunction} a) for different junction lengths, $L'$, while keeping constant its width and the strength of $\lambda_{\rm KM}$.
We see that, with increasing junction length, the polarization effect of the \gls*{qsh} edge states is increased due to the reduced evanescent bulk contribution.

We can further analyze the valley-filtering performance of the junction by combining our numerical results with the analytical ones of Ref.~\cite{Bonasera_2025_a}, obtained using the graphene low-energy description and periodic boundary conditions along the width of the junction.
When using periodic boundary conditions, the transverse momentum is a good quantum number and is conserved during the scattering problem across the junction.
This means that every incoming bulk state is characterized by its own transmission probability, which, at zero chemical potential $\mu=0$, was found to be
\begin{equation}\label{eq:App:Analytical_BulkTransmissions}
    \tau_{\rm Bulk}^{s\nu} \left( k_n \right) = \cosh \left( \frac{L'}{L} \sqrt{ 1 + \left(2 \pi \frac{L}{W'} n \right)^2 } \right)^{-2},
\end{equation}
%
independent of the spin z-projection, $s=\pm$, and valley, $\nu=K,K'$, degrees of freedom; $L'$ and $W'$ are the junction's dimensions, $L = \hbar v /\lambda_{\rm KM}$, and we have considered the infinite doping limit of the pristine leads, $\mu_{\rm R} \to \infty$, with $n = 0, 1\dots, \infty$ defining the allowed transverse momenta $k_n=2\pi n/W'$.
At this point, we can redefine the valley polarization as a weighted sum of fully polarized contributions from bulk states, belonging to a definite valley, and the two transparent helical edge states as
\begin{equation}\label{eq:App:Analytical_ValleyPolarization}
    \kappa\left( \mu =0 \right) = \frac{\sum_{n,s}\tau_{\rm Bulk}^{sK} - \sum_{n,s}\tau_{\rm Bulk}^{sK'} + 2}{\sum_{n,s,\nu}\tau_{\rm Bulk}^{s\nu} + 2}~,
\end{equation}
both in the numerator and denominator, there is a $2$ which arises from the presence of two transparent helical edge states.
Fig.~\ref{fig:App:VP_KM_DoubleJunction} b) shows the results obtained for $\kappa$. 
We observe that, largely regardless of its width $W'$, a \gls*{qsh} junction achieves strong valley polarization, with $\kappa \gtrsim 0.9$, when its length $L'$ is about twice the characteristic length $L = \hbar v/\lambda_{\rm KM}$.
Fig.~\ref{fig:App:VP_KM_DoubleJunction} c) shows $\kappa$ along a vertical cut of panel b) for the value $W'=5L$ that we have used in the numerical calculations: the black dots are the simulated values shown in panel a) for $\mu = 0$, which match quite well with the analytical predictions.

We also note that we can obtain an opposite valley polarization, for the $K'$ valley, by reversing the current direction (for example, by injecting current from the left lead and collecting it into the right one) or by using the opposite doping in the pristine leads, $\mu_{\rm R} \to -\mu_{\rm R}$.
In this case, the edge states would have the opposite contribution in the numerator of Eq.~\eqref{eq:App:Analytical_ValleyPolarization}, namely $2 \to -2$.

\section{valley polarizationlarization in a QVSH single junction}\label{sec:App:QVSH_ValleyPolarization}

\begin{figure*}[t!]
\centering
\begin{overpic}[width=.6\textwidth]{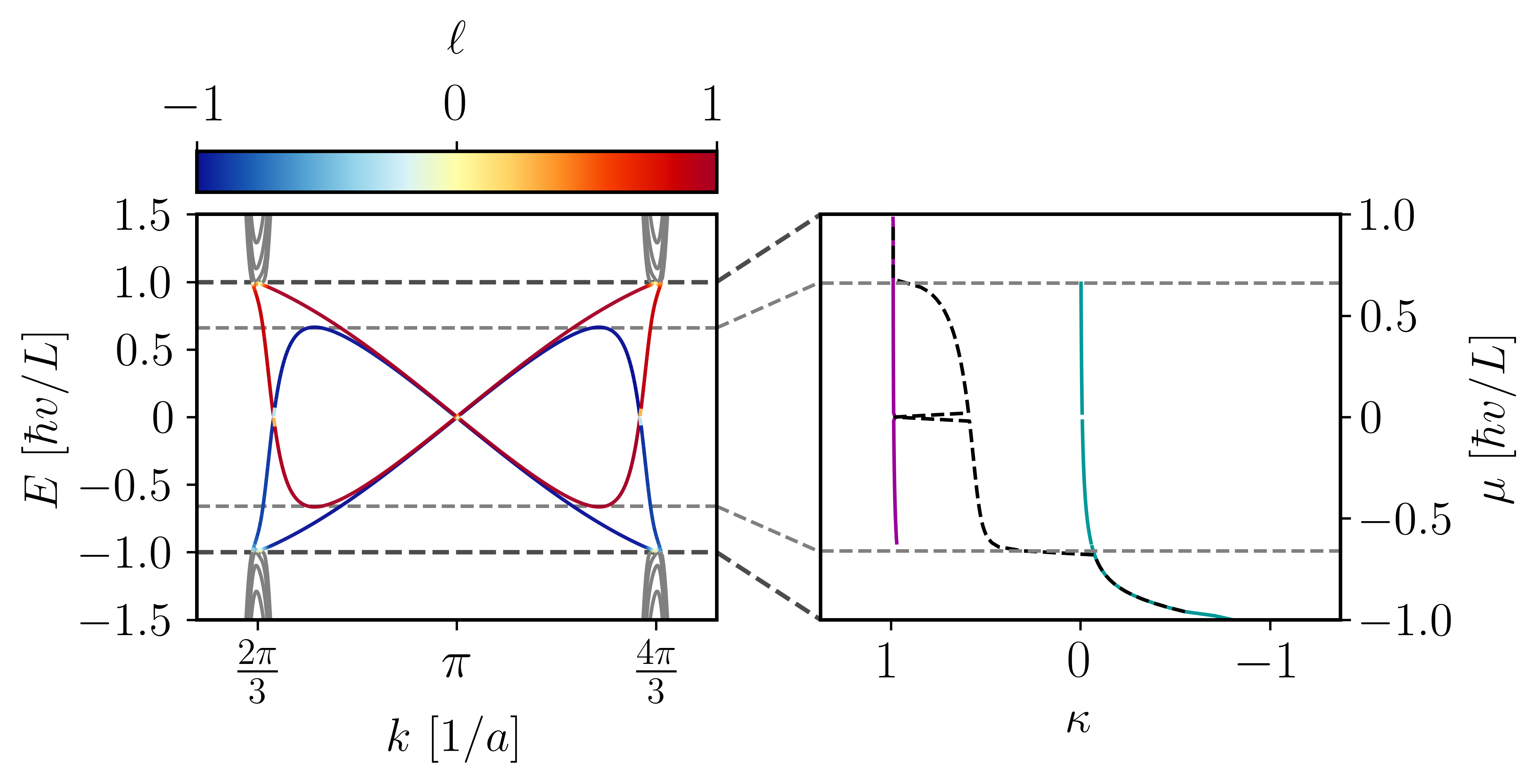}\end{overpic}
\caption{The left panel shows the band structure of a \gls*{qvsh} zigzag graphene ribbon: as in Fig.~\ref{fig:JP_EC_VZ} b), the color scale indicates the degree of edge localization, defined in Eqs.~\eqref{eq:Localization_Definition} and \eqref{eq:Localization_Definition_WeightingDifferent}, in red (blue) for the bottom (top) edge, the dashed black horizontal lines highlight the energy band gap, and the dashed gray horizontal lines define the energy range in which the edge states exist on both edges of the junction; the energies are measured in units of $\hbar v /L$ and the momenta in units of $1/a$.
In the right panel, we show the valley polarization of the incoming current for a single junction between pristine and \gls*{qvsh} zigzag graphene: in magenta (cyan) we show the results for the two transmission eigenstates producing $G_{\rm B}$ ($G_{\rm T}$) of the main text, while in dashed black we show the results extended to all transmission eigenstates.
The $\kappa$ axis is reversed to align with the appearance of the $K = 2\pi/3~[1/a]$ and $K'= 4\pi/3~[1/a]$ valleys in the left panel.
}
\label{fig:App:Valley_Polarization_R5S1}
\end{figure*}

Here, we explore in more detail the valley polarization effect in the single junction between pristine and \gls*{qvsh} zigzag graphene. The dimensions and parameters of the junction are the same as in the main text.
Fig.~\ref{fig:App:Valley_Polarization_R5S1} shows the valley polarization of the incoming current for the mentioned scattering problem.
The valley polarization, $\kappa$, is computed using Eqs.~\eqref{eq:TransmisssionEigenproblem_Original} and \eqref{eq:IncomingValleyPolarization}.
Here, the sum is limited to the two highest (second highest) transmission eigenstates for the magenta (cyan) line, preserving continuity, while it spans all transmission eigenstates for the dashed black one.
The highest (second highest) transmission eigenstates are the ones that match at the interface with the pseudohelical and valley-edge states of the bottom (top) edge to give rise to the conductance $G_{\rm B}$ ($G_{\rm T}$) of the main text.
The only exception is at $\mu = 0$ where only the pseudohelical states exist, while the valley-edge ones become gapped due to finite-size effects, as can be seen from the band structure in the left panel of Fig.~\ref{fig:App:Valley_Polarization_R5S1}.
%
This finite-size effect produces the discontinuity visible in the valley polarization plot.
Moreover, it shows that, as reported in the main text, the \gls*{qvsh} pseudohelical edge states provide a similar level of valley polarizationlarization as the \gls*{qsh} helical one that have been primarily investigated.
For $\mu \neq 0$ the valley-edge states also contribute to transport: the one that matches the valley polarization of the incoming current combines with the pseudohelical state of the same edge to produce two fully valley polarized and high transmission channels, while the other one produces two fully unpolarized low transmission channels on the other edge of the junction, as explained in the main text in Sec.~\ref{sec:QVSH_AsymmetricTransport}.

\section{Energy dispersion for more armchair content}\label{sec:App:EnergyDispersion_MoreArmchair}

\begin{figure*}
\centering
\begin{overpic}[width=.75\textwidth]{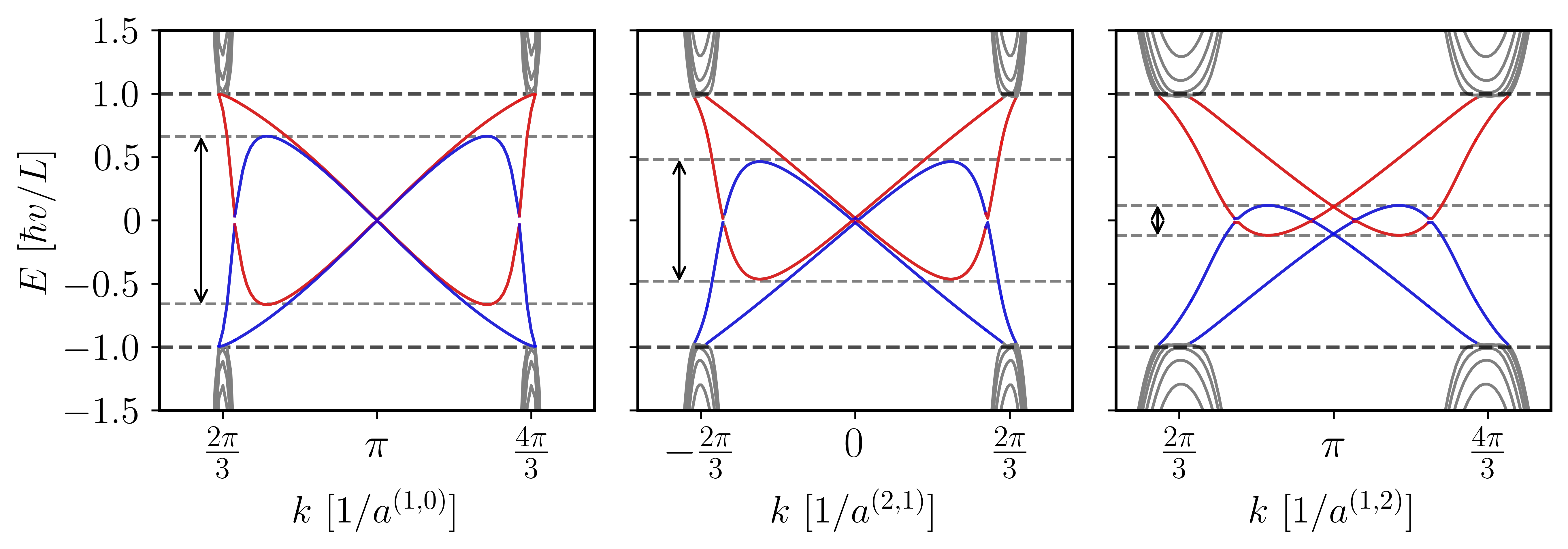}\end{overpic}
\caption{Energy dispersion of a \gls*{qvsh} ribbon with increasing armchair content: from left to right: $(m,n) = (1,0)$, $(m,n) = (2,1)$, $(m,n) = (1,2)$; the energies are measured in units of $\hbar v /L$ and the momenta in units of $1/a^{(m,n)}$.
The edge states are colored red or blue depending on their localization on the bottom or top edge of the ribbon, respectively; differently from Fig.~\ref{fig:JP_EC_VZ} b) or Fig.~\ref{fig:App:Valley_Polarization_R5S1}, there is no color spectrum indicating the degree of localization.
The vertical arrows highlight the energy window where the edge states coexist on both edges of the ribbon.
Note that the primitive cell of the periodic ribbon has a different length in the three cases with $1 = a = a^{(1,0)} < a^{(2,1)} < a^{(1,2)}$.}
\label{fig:App:EnergyDispersion_ArmchairContent}
\end{figure*}

In the main text, we refer to two behaviors of the energy dispersion for \gls*{qvsh} ribbons with increasing armchair content in their edge termination: the energy range in which edge states are present on both sides of the junction becomes smaller, and the dispersion of the edge-state energies becomes progressively steeper.

Fig.~\ref{fig:App:EnergyDispersion_ArmchairContent} shows the energy dispersion for \gls*{qvsh} ribbons with the same characteristics of width and \gls*{soc} parameters as in the main text, but with increasing armchair content in their edge termination; from left to right: $(m,n) = (1,0)$, $(m,n) = (2,1)$, $(m,n) = (1,2)$.
We see that the energy window of coexistence of edge states on both sides of the ribbon, highlighted by the vertical arrows in the plots, gets smaller as the armchair content increases.

Regarding the latter point, we can give an estimate of the edge states velocities as the energy band gap, $ 2\lambda_{\rm VZ}$ with the chosen parameters ($\lambda_{\rm VZ}<\lambda_{\rm R}$), divided by the momentum range spanned by the edge states, $\Lambda^{(m,n)}$, as
\begin{equation}\label{eq:App:Velocities_EdgeStates}
    v_{\rm Edge}^{(m,n)} \sim \frac{2 \lambda_{\rm VZ}}{ \Lambda^{(m,n)}}.
\end{equation}
We compute $\Lambda^{(m,n)}$ for the different edge terminations as
\begin{equation}\label{eq:App:DeltaMomentum_EdgeStates}
    \Lambda^{(m,n)} = \frac{\left[ 3 + \left(-1\right)^m \right]\pi}{3 a^{(m,n)}},
\end{equation}
which coincides with other methods in literature, such as in Ref.~\cite{Delplace_2011_a}.
Using the above two equations we get, as previously stated, increasing velocities with increasing armchair content in the edge termination as:
$v_{\rm Edge}^{(1,2)}/v_{\rm Edge}^{(2,1)} \approx 2.42$  and  $v_{\rm Edge}^{(1,2)}/v_{\rm Edge}^{(1,0)} \approx 4.36 $.
Using Eq.~\eqref{eq:App:DeltaMomentum_EdgeStates}, we can also estimate the total number of constructive interference peaks per edge termination as $n^{(m,n)} \sim \left[ \Lambda^{(m,n)} / (2\pi/L) \right]$, where $\left[ x \right] $ denotes the integral part of $x$. We get $n^{(m,n)}\approx 19,10,4$, for $(m,n)=(1,0),(2,1),(1,2)$, respectively, which roughly matches the number of peaks simulated in Fig.~\ref{fig:Crit_Current_R5S1_CleanAndDisordered} a) of the main text.

\section{Critical current for selected disorder configurations}\label{sec:App:CriticalCurrent_DisorderedConfigurations}

\begin{figure*}[t!]
\centering

\begin{overpic}[width=0.3\textwidth,trim={0 0cm 0 0cm}]{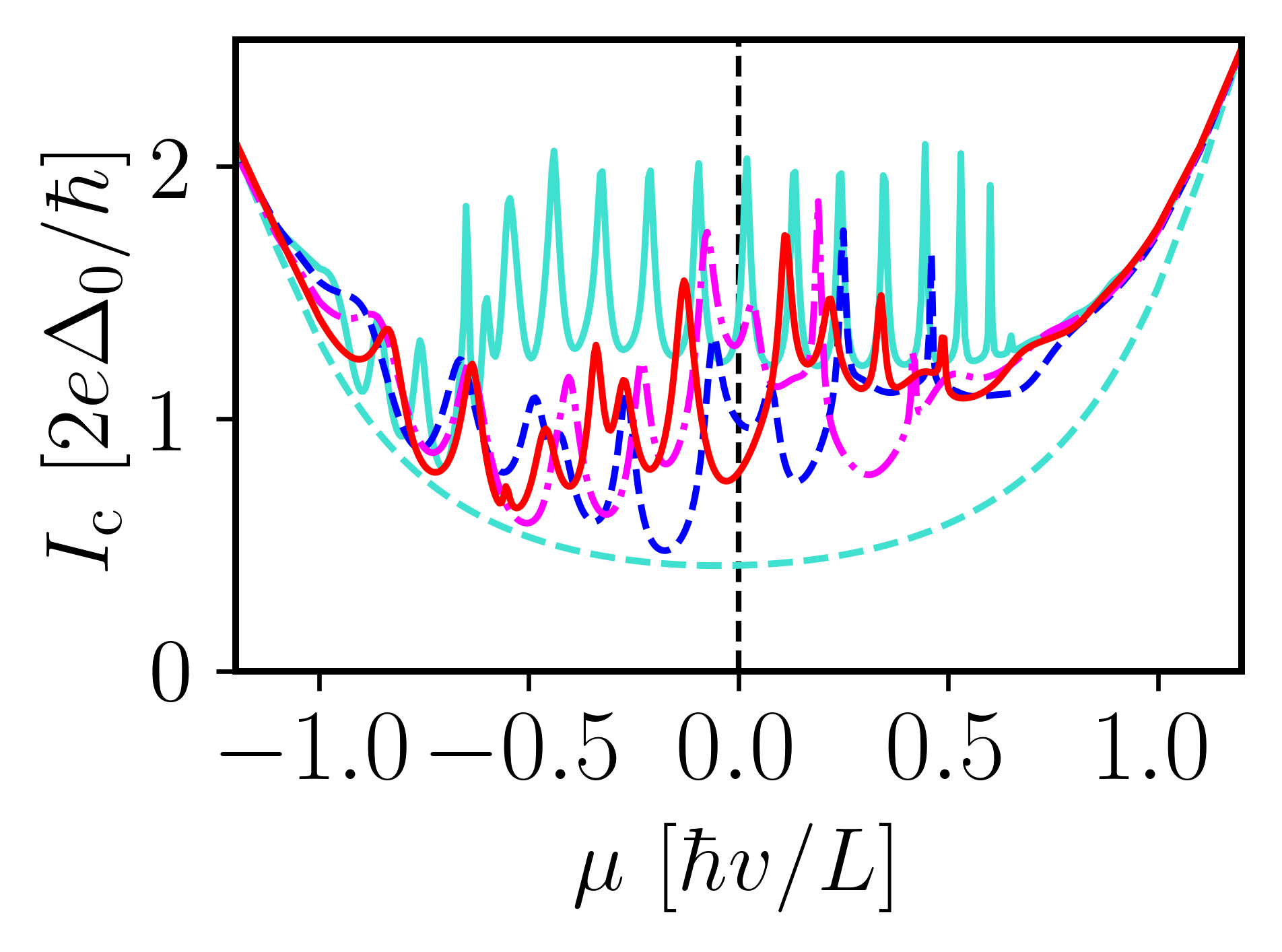}\put(2,70){a)}\end{overpic}
\begin{overpic}[height=0.22\textwidth,trim={0 0cm 0 0cm}]{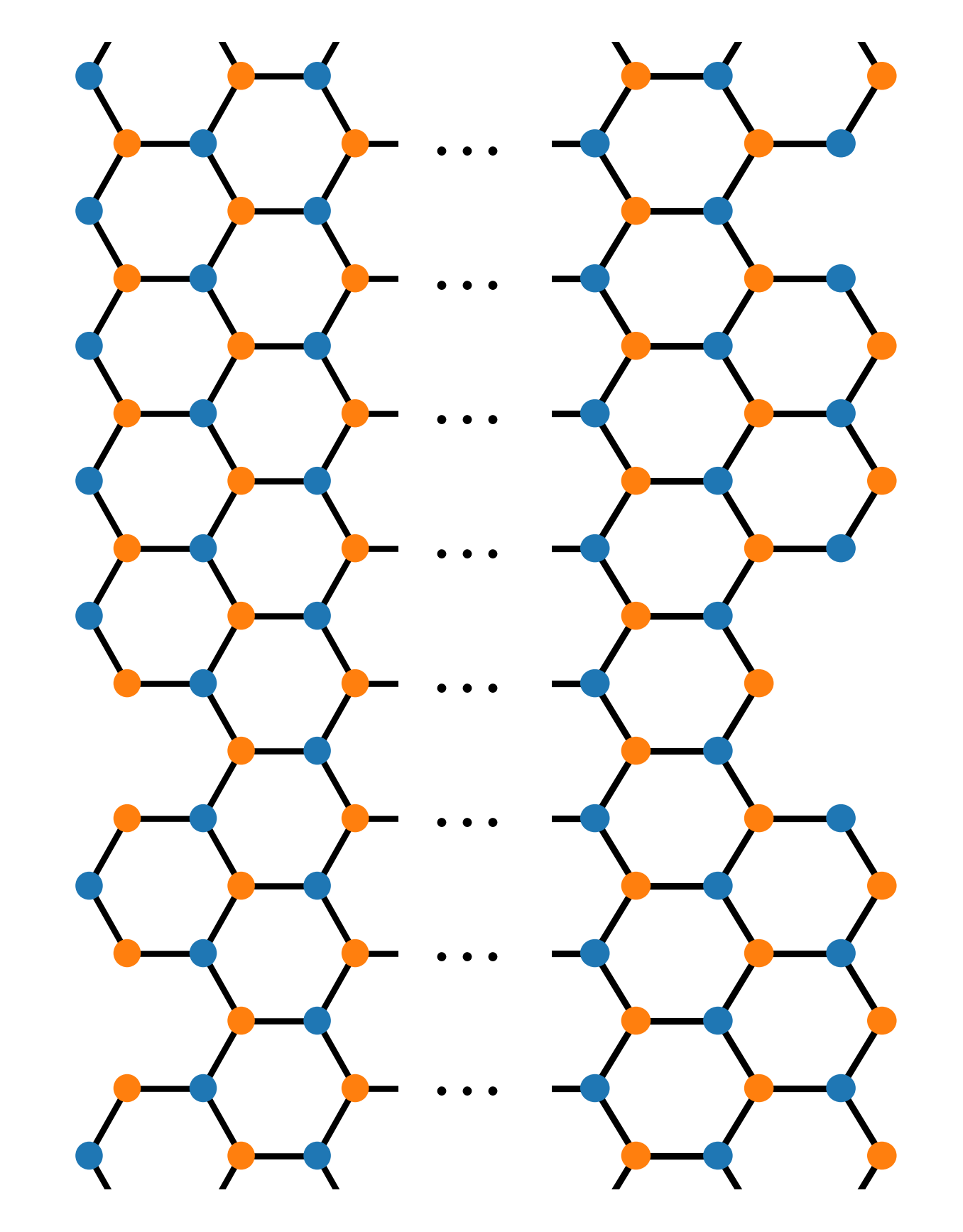}\put(-4,95){b)}\end{overpic}
\begin{overpic}[width=0.3\linewidth]{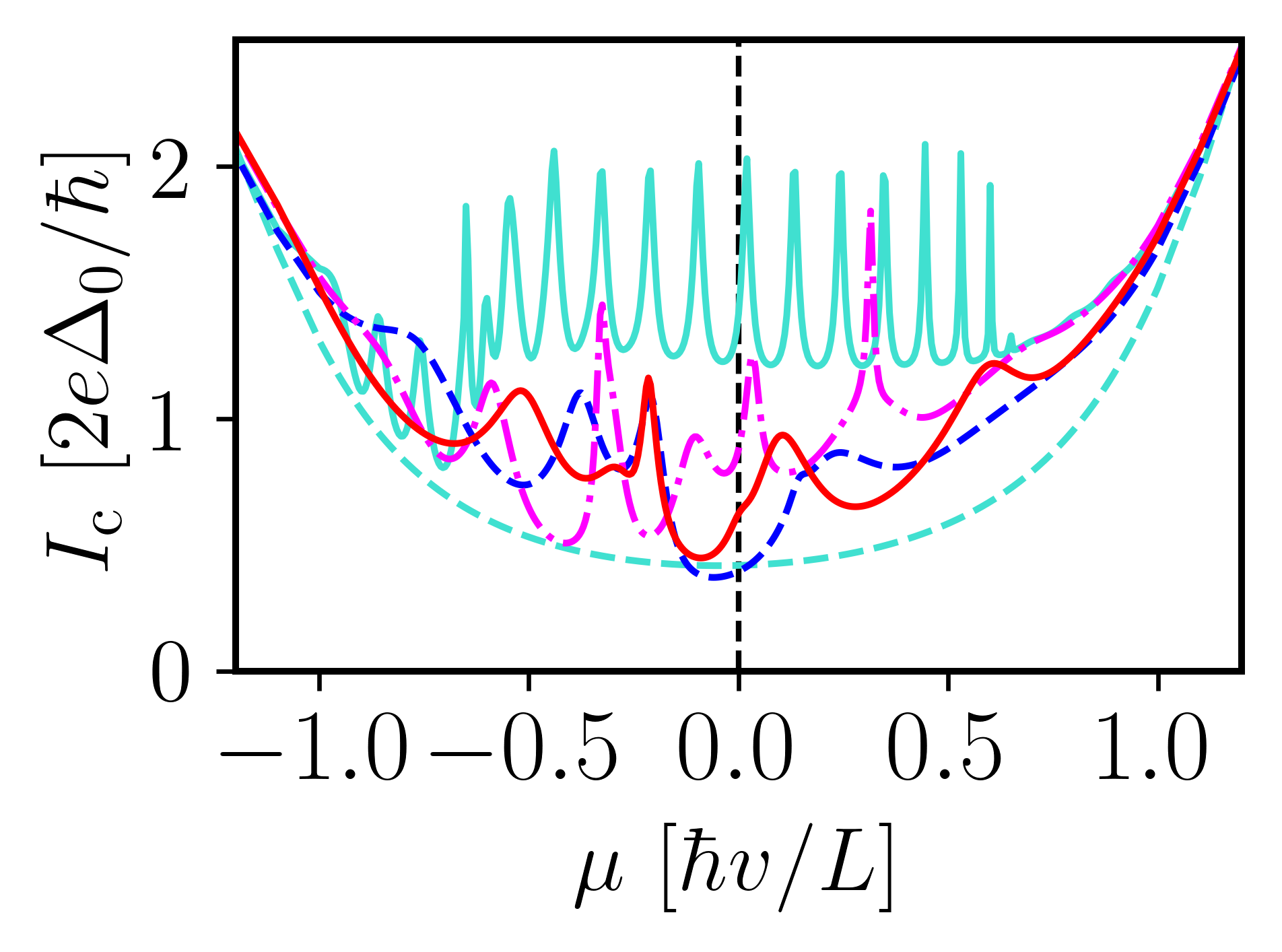}\put(2,70){c)}\end{overpic}
\begin{overpic}[height=0.22\textwidth,trim={0 0cm 0 0cm}]{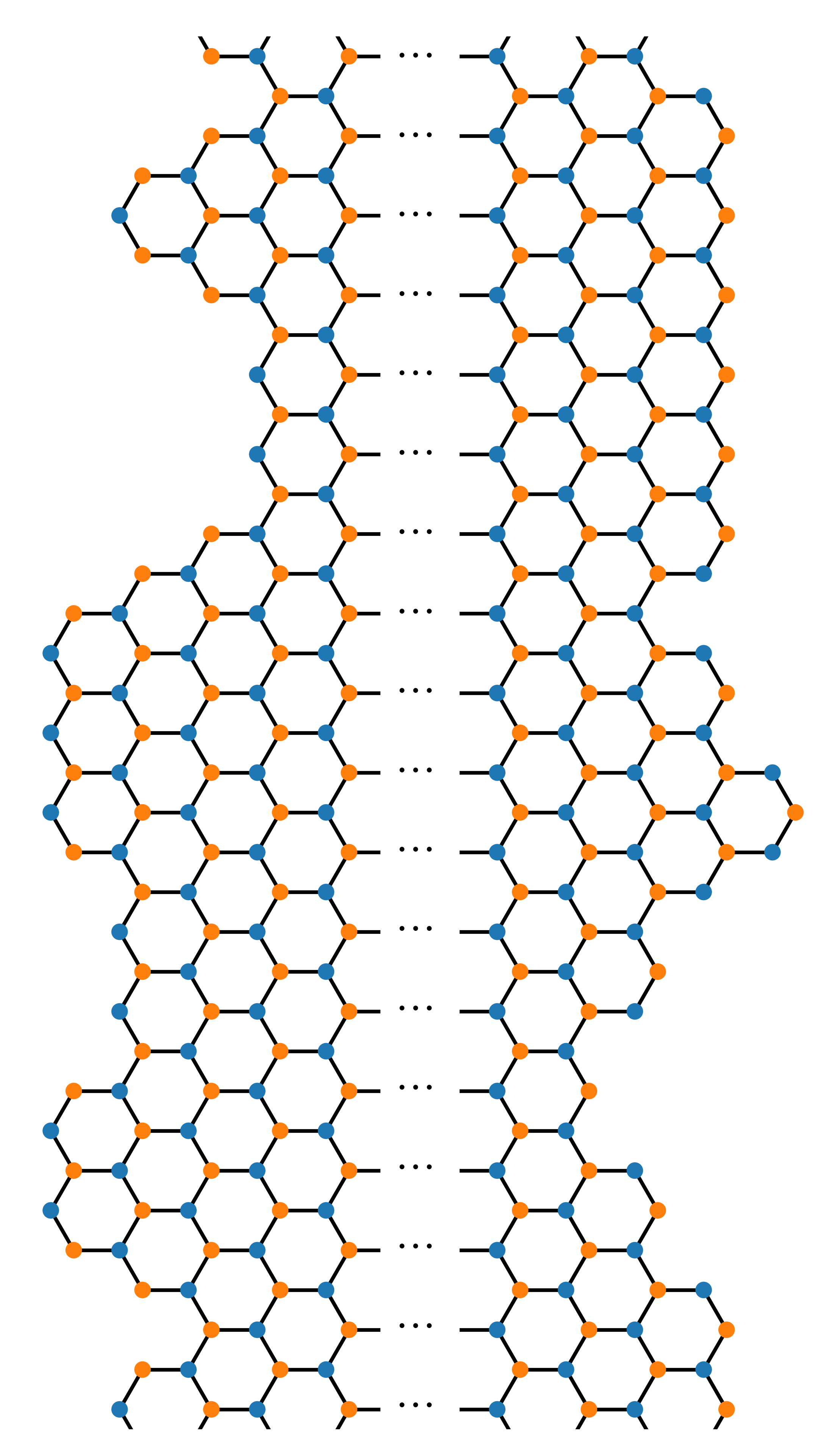}\put(-4,95){d)}\end{overpic}

\caption{Critical current $I_{\rm c}$ (in units of $2e\Delta_0/\hbar$) of a \gls*{gjj}  as a function of the Fermi level $\mu$ (in units of $\hbar v/L$) for a few configurations of small disordered zigzag terminations in a) and heavy disordered zigzag terminations in c); the scattering region is made of \gls*{qvsh} graphene with the same parameters used in the main text, namely $\lambda_{\rm R} = 5~ \hbar v/L$, $\lambda_{\rm VZ} = \hbar v /L$ and a junction length of $L=59a$, and a width $W=5L$ (when non-disordered). As a reference, we show in cyan the results for a clean zigzag termination, in solid, and a clean armchair one, in dashed, taken from Fig.~\ref{fig:Crit_Current_R5S1_CleanAndDisordered}b) of the main text.
We also show an example section of the edges for small and heavy disorder in b) and d), respectively.}
\label{fig:Crit_Current_R5S1_ExampleRuined}
\end{figure*}

Fig.~\ref{fig:Crit_Current_R5S1_ExampleRuined} displays the critical current $I_{\rm c}(\mu)$ as a function of $\mu$ for a \gls*{gjj} with selected small and heavy disordered zigzag terminations in panels a) and c), respectively. For comparison, the results for a clean zigzag (solid cyan) and a clean armchair (dashed cyan) junction are also shown.
Here, b) (d)) represents an example section of zigzag terminations with small (heavy) disorder.
As in the main text, the scattering region is made of graphene in the \gls*{qvsh} phase with $\lambda_{\rm R} = 5~ \hbar v/L$, $\lambda_{\rm VZ} = \hbar v /L$ and a junction length of $L=59a$, and a non-disordered width of $W=5L$.
We find that nearly every disordered configuration exhibits some level of constructive interference, but this effect is obscured by the averaging procedure used to obtain the final results presented in Fig.~\ref{fig:Crit_Current_R5S1_CleanAndDisordered} of the main text.

\section{Spatial distribution of the current for a junction with armchair terminations}\label{sec:App:CleanArmchair_CurrentPlot}

\begin{figure}
\centering
\begin{overpic}[width=0.86\columnwidth]{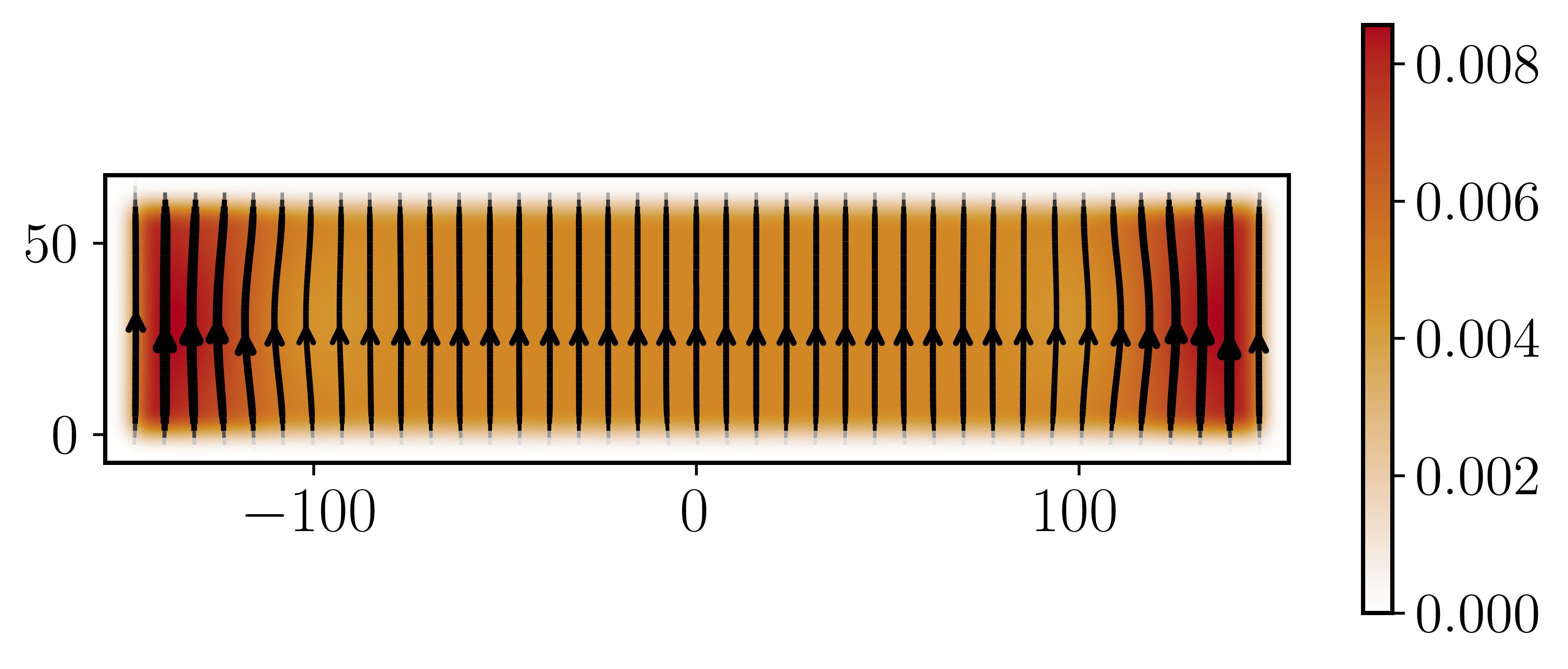}\end{overpic}
\caption{Spatial distribution of the current for a double junction with armchair graphene in the \gls*{qvsh} phase: the junction's width and \gls*{soc} parameters are the same as in the main text, namely $L=59a$ and $W=5L$, $\lambda_{\rm VZ} = \hbar v/L$ and $\lambda_{\rm R} = 5~\hbar v/L$, and the Fermi level is set at $\mu=0$.
For simplicity, the device is rotated $90$ degrees clockwise in the plot.}
\label{fig:App:LocalCurrents_Armchair_R5S1}
\end{figure}

As discussed in the main text, in the short-junction limit and in the absence of a magnetic field, the supercurrent through the junction can be directly related to the transmission channels of a corresponding non-superconducting double junction.
In Fig.~\ref{fig:App:LocalCurrents_Armchair_R5S1}, we show the spatial distribution of the current obtained by solving the scattering problem for a non-superconducting double junction with armchair graphene in the \gls*{qvsh} phase: the junction's width and \gls*{soc} parameters are the same as in the main text, and the Fermi level is set to $\mu=0$.
Although all the transmission channels are evanescent bulk channels, we see that the local distribution of the current is not perfectly homogeneous throughout the junction.
It shows a slight accumulation along the edges, which produces a deviation, both in periodicity and damping, from the typical Fraunhofer pattern discussed in the main text, in Sec.~\ref{sec:SupercurrentRobustness}.

\section{Graphene symmetries and SDE for other edge terminations}\label{sec:App:SDE_NonZigzag}

\begin{figure}[t!]
\centering
\begin{overpic}[width=1\columnwidth]{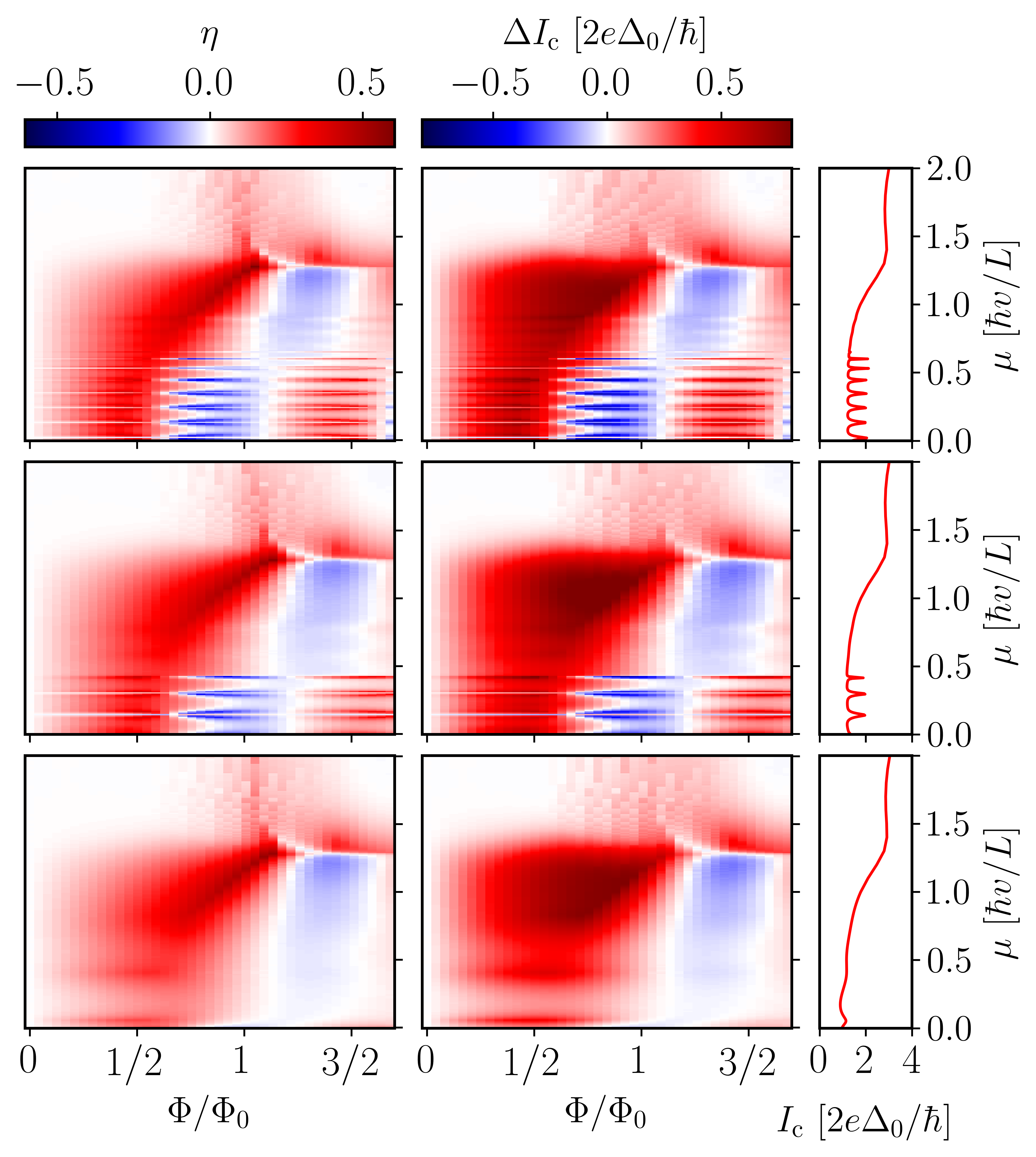}\end{overpic}
\caption{Efficiency, $\eta$, (left column) and non-reciprocal supercurrent, $\Delta I_c$, (middle column) for a \gls*{gjj} in the \gls*{qvsh} phase with $(m,n)=(1,0)$ (first row), $(m,n)=(2,1)$ (second row), $(m,n)=(1,2)$ (third row) terminations, as functions of the Fermi level of the graphene scattering region, $\mu$ (vertical axis), and the magnetic flux, $\Phi$, threading the junction (horizontal axis).
As reference, in the third column we show the critical current (horizontal axis) as a function of the Fermi level (vertical axis) for a junction with the relative termination at zero magnetic flux, the same results shown in Fig.~\ref{fig:Crit_Current_R5S1_CleanAndDisordered}~a). The critical currents are shown in units of $2e\Delta_0/\hbar$, the energies in units of $\hbar v/L$ and the magnetic flux in units of the superconducting flux quantum $\Phi_0 = \hbar/2e$.}
\label{fig:App:SDE_AngledTerminations}
\end{figure}

The \gls*{sde} that we have studied in the main text is due to the orbital effects of the magnetic field (we neglect the Zeeman coupling), and is rooted in the asymmetric current trajectories within the junction with respect to the transverse direction of transport, as obtained also in Ref.~\cite{Chirolli_2025_a}.

To study the symmetry properties of the \gls*{qvsh} graphene region, we first expand the tight-binding Hamiltonian in Eq.~\eqref{eq:Hamiltonian} around the Dirac points to obtain a low-energy description of the system as \cite{CastroNeto_2009_a,Frank_2018_a}
\begin{subequations}\label{eq:App:ContinuumHamiltonian}
\begin{align}
    H & = H_K + H_{\rm S}, \label{eq:App:TotalContinuumHamiltonian} \\
    H_{\rm K} & = \hbar v \left( k_x \tau_z \sigma_x - k_y \sigma_y \right) -\mu, \label{eq:App:KineticContinuumHamiltonian} \\
    H_{\rm S} & = \lambda_{\rm VZ} s_z \tau_z - \lambda_{\rm R} \left( s_y \tau_z \sigma_x + s_x \sigma_y \right), \label{eq:App:SOCContinuumHamiltonian}
\end{align}
\end{subequations}
where $H_{\rm K}$ and $H_{\rm S}$ represent the kinetic and \gls*{soc} terms, respectively.
The Hamiltonian in Eq.~\eqref{eq:App:ContinuumHamiltonian} describes an infinite graphene lattice with the carbon atoms arranged in an armchair (zigzag) pattern along the $x$ ($y$) direction.
The kinetic term of the Hamiltonian is mirror symmetric in both the $x$ and $y$ directions, with respect to the operators
\begin{subequations}
\begin{gather}
    P_x =  - i s_x \sigma_y, \label{eq:App:MirrorXOperator} \\
    P_x H_{\rm K}\left( k_x,k_y \right)P_x^{-1} = H_{\rm K}\left(- k_x,k_y \right),
\end{gather}
\end{subequations}
and
\begin{subequations}
\begin{gather}
    P_y =  i s_y \tau_x \sigma_z, \label{eq:App:MirrorYOperator} \\
    P_y H_{\rm K}\left( k_x,k_y \right)P_y^{-1} = H_{\rm K}\left( k_x,-k_y \right),
\end{gather}
\end{subequations}
respectively.
For a junction with zigzag (armchair) terminations, the $P_x$ ($P_y$) is the mirror operator with respect to the transverse current direction.

After introducing the \gls*{soc} term we find that, for a junction with armchair terminations, the graphene Hamiltonian is still mirror symmetric along the transverse current direction
\begin{equation}\label{eq:App:ExtendedInversionSymmetry}
    P_y H(k_x,k_y) P_y^{-1} = H(k_x,-k_y),
\end{equation}
which means that for each trajectory of the current across the junction, there is a symmetric one, which forbids the orbital \gls*{sde}.
The same cannot be done using the $P_x$ operator for junctions with zigzag edge terminations,
\begin{equation}\label{eq:App:ExtendedInversionNoSymmetry}
    P_x H(k_x,k_y) P_x^{-1} - H(-k_x,k_y)= -2\lambda_{\rm VZ} s_z \tau_z.
\end{equation}
Indeed, as discussed in the main text, junctions with zigzag edge terminations show non-reciprocal superconducting transport.
We observe the same behavior for other junctions with non-armchair edges, which display features analogous to the zigzag case and achieve similar maximum efficiencies, up to $\eta \approx 0.6$. In Fig.~\ref{fig:App:SDE_AngledTerminations}, we show the dependence of $\eta$ and $\Delta I_{\rm c}$ on the threading magnetic flux $\Phi$ and the Fermi level $\mu$ for a \gls*{gjj} hosting graphene in the \gls*{qvsh} phase with several different edge terminations.

\end{document}